\documentclass{aa}
\pdfoutput=1

\usepackage{amsmath,bm} 
\usepackage[varg]{txfonts}
\usepackage{natbib}
\usepackage{chemformula}
\usepackage{afterpage}
\usepackage[markup=bfit]{changes}
\usepackage{xcolor}
\usepackage{pdflscape}
\usepackage{mathtools}

\usepackage[colorlinks=true,linkcolor=blue,citecolor=blue,urlcolor=blue,breaklinks]{hyperref}

\bibpunct{(}{)}{;}{a}{}{,}
\usepackage{booktabs}
\usepackage{gensymb}

\usepackage{natbib,twoopt}
\defcitealias{vanzee06}{VZ06}
\defcitealias{ruizescobedo18}{R18}
\defcitealias{kirby17}{K17}

\usepackage{tablefootnote}

\newcommand{\hb}{H$\beta$ }
\newcommand{\hy}{H$\gamma$ }  
\newcommand{\oiiiauroral}{$[\ion{O}{iii}]\lambda4363$ }
\newcommand{\oiiistrong}{$[\ion{O}{iii}]\lambda5007$ }
\newcommand{\oiiistrongweak}{$[\ion{O}{iii}]\lambda4959$ }
\newcommand{\mets}{$12+\log(\mathrm{O/H})$ }

\newcommand{\oii}{$[\ion{O}{ii}]\lambda3727$ }
\newcommand{\op}{$\mathrm{O^{+}}$ }
\newcommand{\np}{$\mathrm{N^{+}}$} 
\newcommand{\sulfurp}{$\mathrm{S^{+}}$}
\newcommand{\hp}{$\mathrm{H^{+}}$ }
\newcommand{\opp}{$\mathrm{O^{++}}$ }
\newcommand{\hii}{\ion{H}{ii} }

\newcommand{\yeff}{$y_{\mathrm{eff}}$ }

\newcommand{\ophp}{$\mathrm{O^{+}/H^{+}}\ $}
\newcommand{\opphp}{$\mathrm{O^{++}/H^{+}}\ $}

\begin{document} 

   \title{The ionisation structure and chemical history in isolated Hii regions of dwarf galaxies with integral field unit}
    \subtitle{II. The Leo A galaxy\thanks{Based on observations taken under the ESO program ID 079.B-0877(A)}}   
   \author{A. Andrade \inst{\ref{unab-casona},\ref{eso}} 
            \and I. Saviane\inst{\ref{eso}} 
            \and L. Monaco \inst{\ref{unab-concepcion}, \ref{inaf-trieste}}
            \and M. Gullieuszik \inst{\ref{inaf-padova}}
          }

   \institute{Universidad Andres Bello, Facultad de Ciencias Exactas, Departamento de Física y Astronomía - Instituto de Astrofísica, Fernández Concha 700, Las Condes, Santiago, Chile. \label{unab-casona}  \\
   \email{a.andradevalenzuela$@$uandresbello.edu}
    \and
        European Southern Observatory, Alonso de Cordova 3107, Vitacura, Casilla 19001, Santiago de Chile 19, Chile. \label{eso} 
    \and 
        Universidad Andres Bello, Facultad de Ciencias Exactas, Departamento de Física y Astronomía - Instituto de Astrofísica, Autopista
        Concepción-Talcahuano, 7100, Talcahuano, Chile. \label{unab-concepcion} 
    \and 
        INAF – Osservatorio Astronomico di Trieste, Via G. B. Tiepolo 11, 34143 Trieste, Italy \label{inaf-trieste} 
    \and 
        INAF - Osservatorio Astronomico di Padova, Vicolo dell’Osservatorio 5, I-35122 Padova, Italy. \label{inaf-padova}}

   \date{Received XX XX, 2025; accepted XX XX, 2025}
 % \abstract{}{}{}{}{} 
% 5 {} token are mandatory
  \abstract
  {Examining the ionised gas in metal-poor environments is key to understand the physical mechanisms regulating galaxy evolution. However, most of the previous studies of extragalactic \hii regions rely on unresolved observations of gaseous structures.}
  {We study the south-western, spatially resolved \hii region of Leo A, one of the most studied gas-rich isolated galaxies in the Local Group. Using archival VIMOS-IFU/VLT data, we investigate its gaseous structure through optical emission lines to gain insights into the present-day drivers of gas physics in this dIrr, and we place constraints on the chemical evolution scenario responsible for its low chemical enrichment.}
  {We mapped the \hb and \oiiistrong flux distributions of the \hii region, fully covered within the $27''\times 27''$ VIMOS field of view. Oxygen abundances were derived with the $T_{e}$-sensitive method, using the auroral \oiiiauroral emission line detection, obtained by integrating spectral fibres of the data cube.}
  {The emission line maps reveals that the strongest emission comes from the south-west region. Differences between the \hp and \opp distributions indicate a stratified distribution of ionic species, likely powered by the young star cluster at the nebular centre. HST/ACS photometry shows that the brightest star ($\sim15M_{\odot}$) is in the centre of both the \hii region and the young star cluster. Photoionisation production rates derived indicate that this star is able to sustain most of the ionisation budget to power the \hii region, although subject to the assumed electron density. We derive an oxygen abundance of $12+\log(\mathrm{O/H})=7.29\pm0.06$ dex, increasing to $7.46\pm0.09$ dex after correcting for temperature fluctuations. These values place Leo A on the low-mass end of the mass-metallicity relation. Chemical evolution models indicate that, under constant accretion, the stellar mass growth and metal enrichment over the last 10 Gyr are successfully reproduced by both the gas-regulator and leaky-box models.}
  {The distribution of young stars in this \hii region reveals similar features to those of the \hii region in the Sagittarius dIrr (SagDIG), supporting a picture in which the present-day evolution of Leo A is dominated by stellar feedback processes, associated with young stars in the cluster ionising the \hii region studied in this work. The combination of mass loss mechanisms and accretion events efficiently reproduces its chemical evolution, suggesting Leo A has evolved under a gas equilibrium regime across its lifetime.}
    %{}


   \maketitle

\section{Introduction}
\nolinenumbers
Dwarf galaxies are the most abundant systems in the Universe \citep{schechter76}, characterised by low stellar masses, luminosities, and rotational velocities \citep{tolstoy09, sanchesjanssen13}. They are commonly classified by their morphologies, as dwarf spheroidals (dSphs), dwarf irregulars (dIrrs), or transition types (dTs; \citealt{tolstoy09}). Additional classification is given by their star-formation histories (SFHs) into “slow” and “fast” dwarfs \citep{gallart15}, or “single” and “two-component” systems \citep{benitez15}. Within the $\Lambda$CMD cosmological framework, dwarfs are considered the building blocks of the hierarchical growth of galaxies \citep{press74}. On the other hand, since stellar mass and metal content evolve with time, their low gas-phase abundances likely resemble the primordial conditions in the early universe \citep{izotov04}, under the open question of whether dIrrs are unevolved systems or not.\\
\\
In the Local Group, dwarf galaxies have been the preferred laboratory to explore galaxy evolution, by long-term observational campaigns coming from space-based observations \citep{bernard09, gallart15, weisz23}, given the ability of HST and JWST to resolve their stellar populations below the old main-sequence turn-off (oMSTO). Deep colour-magnitude diagrams (CMDs) of dwarf galaxies revealed the presence of old stellar populations \citep{tolstoy09, monelli10}. For dIrrs, star formation episodes across cosmic time have been revealed by red clump and blue plume features in their CMDs \citep{weisz14}, giving insights into diverse SFHs \citep{weisz11}, likely shaped by reionisation, stellar feedback, accretion, and environmental processes, with relative importance of those mechanisms under a galaxy-by-galaxy basis (\citealt{mcquinn24b} and references therein).\\
\\
Gas-phase metallicities are lower than typical star-forming galaxies ($>10^{9}M_{\odot}$), and their position at the low-mass end of the mass-metallicity relation (MZR) highlights a shallower slope and high scatter  \citep{lee06, saviane08, berg12, zahid12}. This behaviour is interpreted by efficient gas removal by either energy or momentum-driven SN winds acting in shallower potential wells  \citep{finlatordave08, dave12, guo16}, supported under analytic chemical evolution models, where the yield\footnote{The yield ($y_{i}$) is defined as the mass fraction of a element $i$ produced by a generation of stars, relative to the fraction of mass locked up in stellar remnants and long lived stars \citep{matteucci21}} shows a positive correlation with stellar mass at the low mass regime \citep{garnet02, tremonti04, chisholm18, tortora22}. However, accretion models without outflows can also reproduce this trend as well as the MZR, as a result of inefficient star formation regulated by the Kennicutt-Schmidt law subject to a critical density threshold \citep{tassis08}\\
\\
Spectroscopic studies of extragalactic \hii regions in dwarfs have been characterised by long-slit observations (e.g, \citealt{saviane02, lee05, vanzee06, skillman13}), with tentative evidence that several nebulae are ionised by single massive OB stars, as two of the four \hii regions in Leo A \citep{gull22}, and the only one in Leo P \citep{telford22}. 

Integral field spectroscopy has provided insights on the global properties of the ionised component in blue compact dwarfs, merger systems, dIrrs, and disk galaxies (e.g. \citealt{james10, james13, james20, perez-montero11, vanzi11, kumari17,kumari18, emsellem22}). However, detailed spatially-resolved studies in individual extragalactic \hii regions remains rare. In this context, our previous work in the only known \hii region of the Sagittarius dIrr (SagDIG) revealed a stratified distribution of ionised species \citep{osterbrock06} likely due to stellar feedback mechanisms \citep{andrade25}, similar to those detected in Galactic and Magellanic \hii regions (e.g., \citealt{sanchez13, barman22, kreckel24}).

For this reason, in this work we perform a similar study to that of \citet{andrade25}, but in one of the four known \hii regions of Leo A (also known as DDO 69 and Leo III).\\
\\
Leo A  is an isolated dIrr at a distance of $\sim800$ Kpc \citep{dolphin02}, with stellar mass of $3.3\pm0.7\times 10^{6}M_{\odot}$ \citep{mcconnachie12, kirby17} and gas mass of $6.9\pm0.7\times10^{6}M_{\odot}$ \citep{hunter12}, known as a gas-rich isolated dIrr in the Local Group. HST observations revealed a delayed (or late-blooming) SFH \citep{cole07, cole14}, decomposed in three phases: (i) an ancient star formation episode ($>$10 Gyr ago), followed by (ii) $\sim$2 Gyr of quiescence, and (iii) an extended (late-blooming) star formation episode beginning $\sim8$ Gyr ago that continues to the present-day. Similar SFHs are reported in Aquarius \citep{cole14}, Leo P \citep{mcquinn24a}, and WLM \citep{mcquinn24b}.\\
\\
Some young main-sequence (MS) massive stars in Leo A, not associated with \hii regions, show evidence of stellar activity, binary dynamics, mass loss, and accretion \citep{gull22}. In contrast, those detected in \hii regions are consistent with expectations from nebular studies \citep{vanzee06, ruizescobedo18, gull22}. In particular, two O-type stars with estimated masses ranging $8 - 30M_{\odot}$ are able to ionise, individually, two eastern \hii regions in Leo A \citep{gull22}. In addition, low-mass young star cluster candidates have been identified in this dIrr, where one of them, labelled C2, lies within the \hii region analysed in this work \citep{stonkute19}.\\
\\
Nebular studies of Leo A presents the four \hii regions of Leo A with similar gas-phase metallicities, as \mets$\sim7.40$ dex ($\sim0.3Z_{\odot}$;  \citealt{vanzee06,ruizescobedo18}). On the other hand, RGB stars show a mean stellar metallicity of $\left< [\mathrm{Fe/H}] \right> = -1.67^{+0.009}_{-0.008}$ dex, where the metallicity distribution is consistent with a pre-enriched closed-box or an accretion scenario \citep{kirby17}, suggesting that Leo A acquired a significant amount of external gas either at early times, or across its lifetime.\\
\\
In this work, we present a detailed analysis of the south-western \hii region using archival VIMOS-IFU/VLT data (Figure \ref{fig0}). We aim to (i) explore the nebular structure of the \hii region, to gain insights about their physical drivers by comparing the young stellar populations with the gaseous structure, (ii) improve the $T_{e}$-based total oxygen abundance estimates to (iii) place constraints on the chemical evolution scenarios tracing the Leo A history from 10 Gyr ago to the present day.\\
\\
The paper is structured as follows: the VIMOS-IFU observations and data reduction are described in Section 2. The emission line maps generated and metallicity estimates based on $T_{e}$ derivations are presented in section 3. In section 4, we explore possible ionisation mechanisms by comparing the distribution of young stars and the location of C2 with the \hii region flux distributions. We also constrain the evolution of Leo A in the last 10 Gyr by using chemical evolution models. We present our conclusions in Section 5.\\
\\
We adopted solar metallicities from \citet{asplund09}, as solar oxygen abundance of $\mathrm{\log(O/H)_{\odot}} = 8.69$ dex and the solar metallicity of $Z_{\odot}=0.0142$.

% ---------------------------------------------------------------------------

\section{Observations and data reduction}
\subsection{VIMOS observations and data reduction}

\begin{figure}
\centering
\includegraphics[width=\hsize]{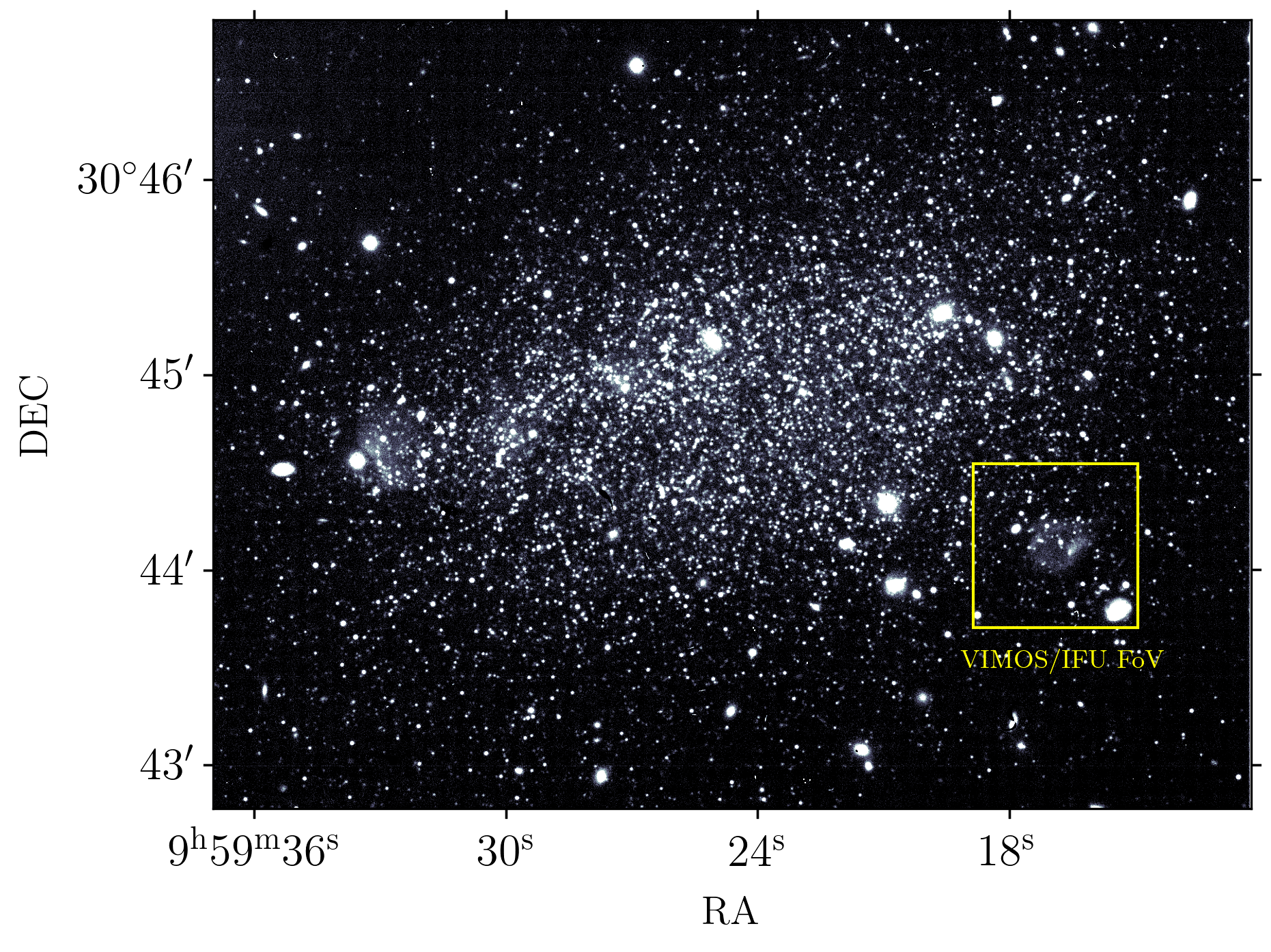}
\caption{Subaru Suprime-Cam H$\alpha$ frame from \citeauthor{stonkute14} (\citeyear{stonkute14}, \citeyear{stonkute19}). The \hii region studied in this work is shown inside the yellow square, representing the VIMOS-IFU FoV.}
\label{fig0}
\end{figure}

Archival data correspond to a single gaseous nebula located in the South West region of Leo A ($\alpha=09^{h}59^{m}17.45^{s},\ \delta=+30\degree 44' 0.5''$; J2000), shown with the yellow square in Figure \ref{fig0}, representing the VIMOS-IFU Field of View (FoV). This \hii region has been previously identified as -101-052 \citep{vanzee06} and \ion{H}{ii}-west \citep{ruizescobedo18}. Observations were obtained with the VIsible MultiObject Spectrograph (VIMOS, \citeauthor{lefevre} \citeyear{lefevre}) under the program 079.B-0877(A) in May 2007 (PI: M. Gullieuszik). VIMOS was a visible (360 nm to 1000 nm) wide-field imager and multi-object spectrograph mounted on Nasmyth focus B of VLT/UT 3 (Melipal). The Integral Field Unit (IFU) mode was used. Two observing blocks (OBs) of one hour each were acquired.  \\
\\
The IFU comprises 1600 fibres (pseudo-slits), where 400 fibres are stored per quadrant. This provides a $27'' \times 27''$ field of view with a spatial resolution of $0.67''$ px$^{-1}$ in the wide-field mode, which is enough to cover the gaseous structure plus adjacent galaxy field free from nebular emission. Therefore, a significant number of fibres were used to generate the sky spectrum, and perform decontamination by telluric lines. The selected spectral setup uses the HR-blue grism ($4150\AA-6200\AA$, $\Delta\lambda = 0.51\AA$ px$^{-1}$, $R = 2020$). Emission lines with S/N $>3$ were detected, including the Balmer emission lines H$\beta$ and H$\gamma$ and the collisional excitation $[\ion{O}{iii}]\lambda\lambda 4959,5007$ emission line doublet.  \\
\\
Data cubes were reduced with the VIMOS Pipeline in the EsoReflex environment \citep{esoreflex}, applying bias subtraction, flat normalisation, wavelength calibration, and flux calibration. The latter was based on spectrophotometric standard stars from \citeauthor{hamuy92} (\citeyear{hamuy92}, \citeyear{hamuy94}) and \citet{moehler14}: The F-type LTT3864 ($V = 12.17, \ B-V =0.50 $), the G-type LTT7379 ($V = 10.23, \ B-V = 0.61$), and the DA-type EG247 ($V = 11.03,\  B-V = -0.14$). The final products are flux-calibrated in units of $10^{-16}\mathrm{erg\ s^{-1}\ cm^{-2}\ \AA^{-1}}$.\\
\\
To reduce the noise in each spectral fibre, a smoothing technique was applied on each of them by weighting each bin by the mean of its three nearest neighbours. Finally, by fitting a second-order Chebyshev polynomial using the Specutils Python library \citep{specutils}, we subtracted the continuum contribution. 

\subsection{Detection of the $[\ion{O}{iii}]\lambda4363$ emission line}
\label{detection}
We aim to estimate the electron temperature, $T_{e}$, for two main reasons: (i) characterise the physical properties of the \hii region, and (ii) derive $T_{e}$-based total oxygen abundances. Regarding the latter, we revisited the total oxygen abundance calculations on this \hii region of \citetalias{vanzee06} and \citetalias{ruizescobedo18}, because the \oii emission line flux measurement of the latter appears to be affected by spectral noise, potentially introducing instrumental biases into their \ophp determinations. \citetalias{vanzee06}, on the other hand, reported metallicities based on a spectrum with a well-detected \oii line, but their results rely on empirical and semi-empirical calibrations rather than direct $T_{e}$ measurements.

\begin{figure*}
\includegraphics[width=\hsize]{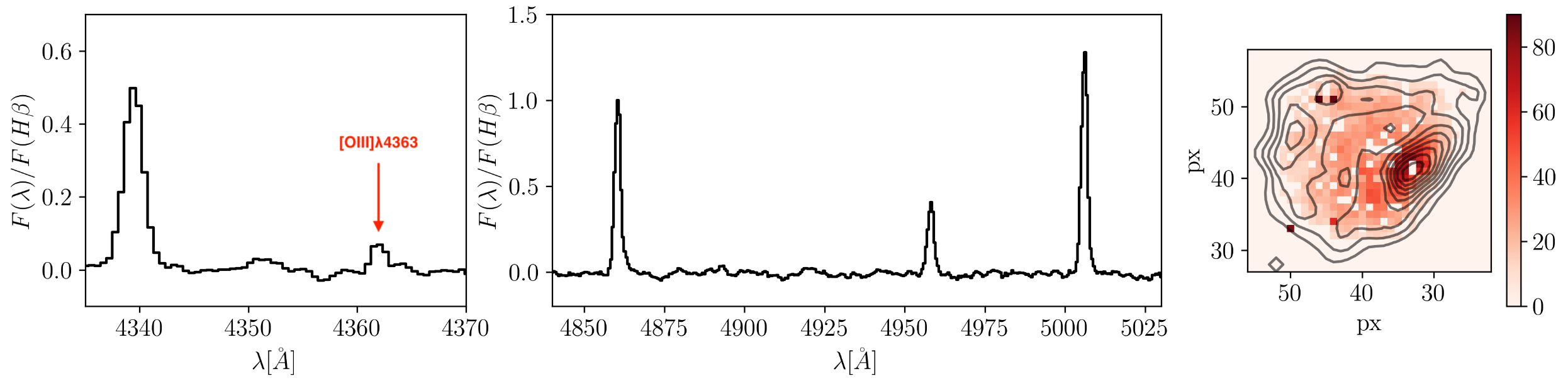}
\caption{Integrated VIMOS-IFU spectrum of the Leo A \hii region normalised by the flux at the peak of the \hb line. Left panel: spectral window showing H$\gamma$ and \oiiiauroral detection, from left to right, respectively. Middle panel: spectral window presenting the H$\beta$, and the $[\ion{O}{iii}]\lambda\lambda 4959,5007$ detection, from left to right, respectively. Right panel: spatial distribution of fibres selected to generate the integrated spectrum of the Leo \hii region with \oiiiauroral detection. The colour code represents the jump value of each selected fibre. The grey contours represent the \hb emission of the nebula as a reference.}
\label{fig1}
\end{figure*}

The procedure makes use of the auroral $[\ion{O}{iii}]\lambda 4363$ emission line (\citealt{peimbert67}, \citealt{aller84}). However, the integrated spectrum using all VIMOS-IFU fibres did not exhibit the auroral line with a signal-to-noise ratio (S/N) greater than 3\footnote{We defined signal (S) as the flux value at the peak of the evaluated emission line, and noise (N) as the standard deviation evaluated in both sides of the line inside a spectral window of 5 px per region.}. For this reason, we applied the \citet{andrade25} procedure to select the fibres which reproduce a detectable (S/N$>3$) auroral line in the integrated spectra.\\
\\
Since auroral and nebular transitions emerge from the same ionisation state (\opp), their spatial distribution is expected to be similar. Therefore, we applied the "jump criterion" in the \oiiistrong emission line. The jump is defined as the ratio between (i) the flux at the peak of the \oiiistrong emission line and (ii) the semi-quartile range of the spectral noise in the $4970-5040\AA$ interval. The semi-quartile range was chosen as it is less sensitive to outliers (i.e. emission lines are outliers with respect to the flux noise). By integrating all those fibres with jump=1, the S/N of the auroral line was calculated. If the S/N is less than 3, we increase the jump threshold and repeat the iteration. At jump=9, the integrated spectrum exhibits a clear \oiiiauroral emission line detection, as the result of combining 398 spectral fibres.\\
\\
Figure \ref{fig1} presents the integrated spectrum of the \hii region. The flux scale is normalised by the flux at the peak of the \hb line. The right panel shows the spatial distribution of the selected fibres, colour-coded by their jump values. The \hb emission line map is superimposed for reference with black contours. Most of the fibres producing the \oiiiauroral detection were found in the zone with most ionisation, at the south-west of the nebula (see also Figure \ref{fig2}). Empty spaxels correspond to fibres lower than the jump threshold.
\\
\\
To explore the chemical evolution of Leo A via gas-phase analysis, it is key to estimate its gas-phase $T_{e}$-based oxygen abundance. However, the VIMOS spectral range does not include the \oii emission line, which prevents us from getting a direct estimate of \ophp. To deal with this limitation, we reproduced the long-slit observations of \citet[hereafter VZ06]{vanzee06} and \citet[hereafter R18]{ruizescobedo18}. This approach allows us to (i) use their \oii fluxes to compute total oxygen abundances and (ii) to probe $T_{e}$ fluctuations across the nebula. The procedure is described in Appendix \ref{appendix:mock_slits}.\\
\\
Emission lines of the integrated spectrum of the \hii region, as well as the VZ06 VIMOS and the R18 VIMOS mock slits, were fitted using single Gaussian curves with the Astropy Modelling library \citep{astropy}. Fluxes were obtained by integrating the Gaussian curves with the Specutils library \citep{specutils}, and uncertainties were derived from the covariance matrix of the fitting parameters.\\
\\
Fluxes were corrected for dust attenuation via Balmer decrement. Assuming Case B recombination ($T_{e}\simeq10^{4}$ K, and $n_{e}=100$ cm$^{-3}$), we adopted the theoretical ratio $I_{\mathrm{H\beta}}/I_{\mathrm{H\gamma}}=$ 2.137 \citep{hummer87}. The reddening constant was calculated as $C_{H\gamma}= [\log(I_{H\beta}/I_{H\gamma})-\log(F_{H\beta}/F_{H\gamma})]/[f(H\beta)- f(H\gamma)]$, where $f(\lambda) = \langle A(\lambda)/A(V) \rangle$ is the extinction law at a given wavelength. The corrected fluxes were then obtained as $I_{\lambda}/I_{H\gamma} = (F_{\lambda}/F_{H\gamma}) \times 10^{C_{H\gamma}[f(\lambda)-f(H\gamma)]}$. Following \citet{cardelli89}, we adopted $R_{V} = 3.1$. The dust-corrected emission line fluxes, normalised to the H$\gamma$ flux, are reported for the integrated spectrum and the mock slits in Table \ref{table1}.

\begin{table*}[]
\caption{Emission line flux measurements on the Leo A \ion{H}{ii} region integrated, the VZ06 mock slit, and the R18 mock slit spectra.}

\begin{tabular}{cccccccc}
\hline\hline
Ion   &  $\lambda[\AA]$    & $F_{\lambda}/F_{H\gamma}$           &  $I_{\lambda}/I_{H\gamma}$ & $F_{\lambda}/F_{H\gamma}$        & $I_{\lambda}/I_{H\gamma}$  & $F_{\lambda}/F_{H\gamma}$            & $I_{\lambda}/I_{H\gamma}$  \\
   &      & VIMOS integrared           &    & VZ06 mock slit        &   & R18 mock slit            &   \\
\hline
H$y$    & 4340 & $1.000 \pm 0.012$                   & $1.000 \pm 0.001$ & $1.000 \pm 0.004$              & $1.000 \pm 0.001$   & $1.000 \pm 0.004$                   & $1.000 \pm 0.001$    \\
$[\ion{O}{iii}]$  & 4363 & $0.124 \pm 0.018$                   & $0.123 \pm 0.018$ & $0.262 \pm 0.003$              & $0.278 \pm 0.051$   & $0.181 \pm 0.003$                   & $0.185 \pm 0.043$    \\
H$\beta$    & 4861 & $2.538 \pm 0.029$                   & $2.137 \pm 0.001$ & $1.967 \pm 0.005$              & $2.137 \pm 0.001$   & $1.722 \pm 0.001$                   & $2.317\pm 0.001$     \\
$[\ion{O}{iii}]$  & 4959 & $1.358 \pm 0.017$                   & $1.112 \pm 0.021$ & $1.098 \pm 0.002$              & $1.149 \pm 0.054$   & $0.986\pm 0.003$                    & $1.267 \pm 0.161$    \\
$[\ion{O}{iii}]$  & 5007 & $3.090 \pm 0.059$                   & $2.949 \pm 0.063$ & $2.819 \pm 0.007$              & $2.951 \pm 0.161$   & $2.306\pm 0.009$                    & $3.031 \pm 0.161$    \\
\hline
$F(H_{\gamma})\times 10^{-16}$ &   --   & $0.063\pm0.002$ &       --            & $0.061\pm0.004$ &       --              & $0.072\pm 0.004$ &          --            \\
$C(H_{\gamma})$ &  --    & $0.132 \pm 0.020$                   &         --          & $0.030 \pm 0.026$              &      --               & $0.004 \pm 0.045$                   &        --              \\
\hline
\end{tabular}

\tablefoot{The first and second columns show the emission lines at their respective rest frame wavelength. The third and fourth column shows the flux measurements and their dust-corrected emission line measurements, normalised by H$\gamma$ flux for the integrated spectrum of the Leo A \hii region, respectively. The fifth and sixth columns are the same for the VZ06 mock slit in the VIMOS-IFU datacube. The seventh and eighth columns are the same for the R18 mock slit in the VIMOS-IFU datacube. The last two rows present the flux measurement of H$\gamma$ and C(H$\gamma$) extinction coefficient, respectively.}
\label{table1}
\end{table*}

%--------------------------------------------------------------------
\section{Data analysis and results}
\subsection{Spatially-resolved flux distributions}
\label{emissionmaps}

\begin{figure*}
\centering
\includegraphics[width=\hsize]{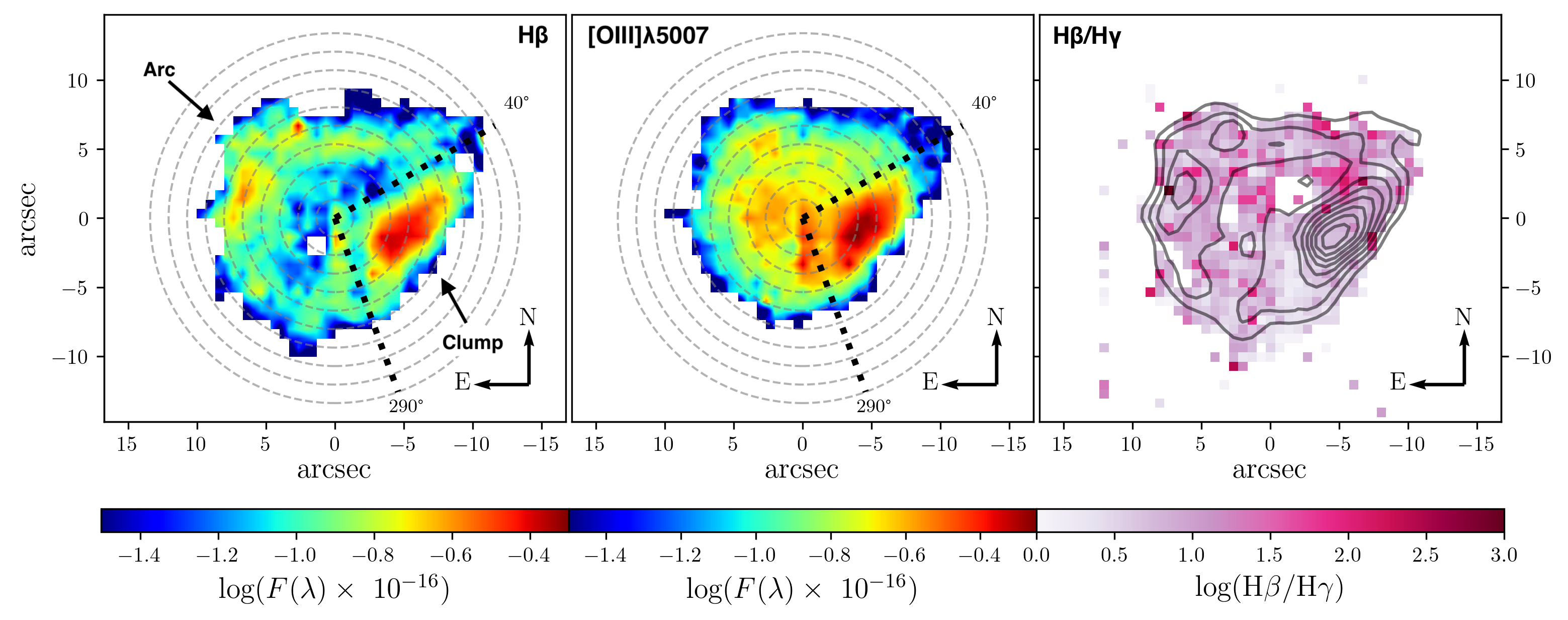}
\caption{Emission line maps of the Leo A \hii region. Left panel: H$\beta$ emission line map. Middle panel: \oiiistrong emission line map. Right panel: H$\beta/$H$\gamma$ map, where the H$\beta$ map contours are superimposed as reference. In all panels, the colour code represents the flux of the emission lines per spectral fibre acquired by fitting Gaussian curves. The grey dashed lines are circles of increasing radius of $1.34''$ (2 px) units up to $13.4''$ (20 px). Black dotted lines show the angles that separate the south-west emission clump and the extended arc.}
\label{fig2}
\end{figure*}

\begin{figure}
\centering
\includegraphics[width=\hsize]{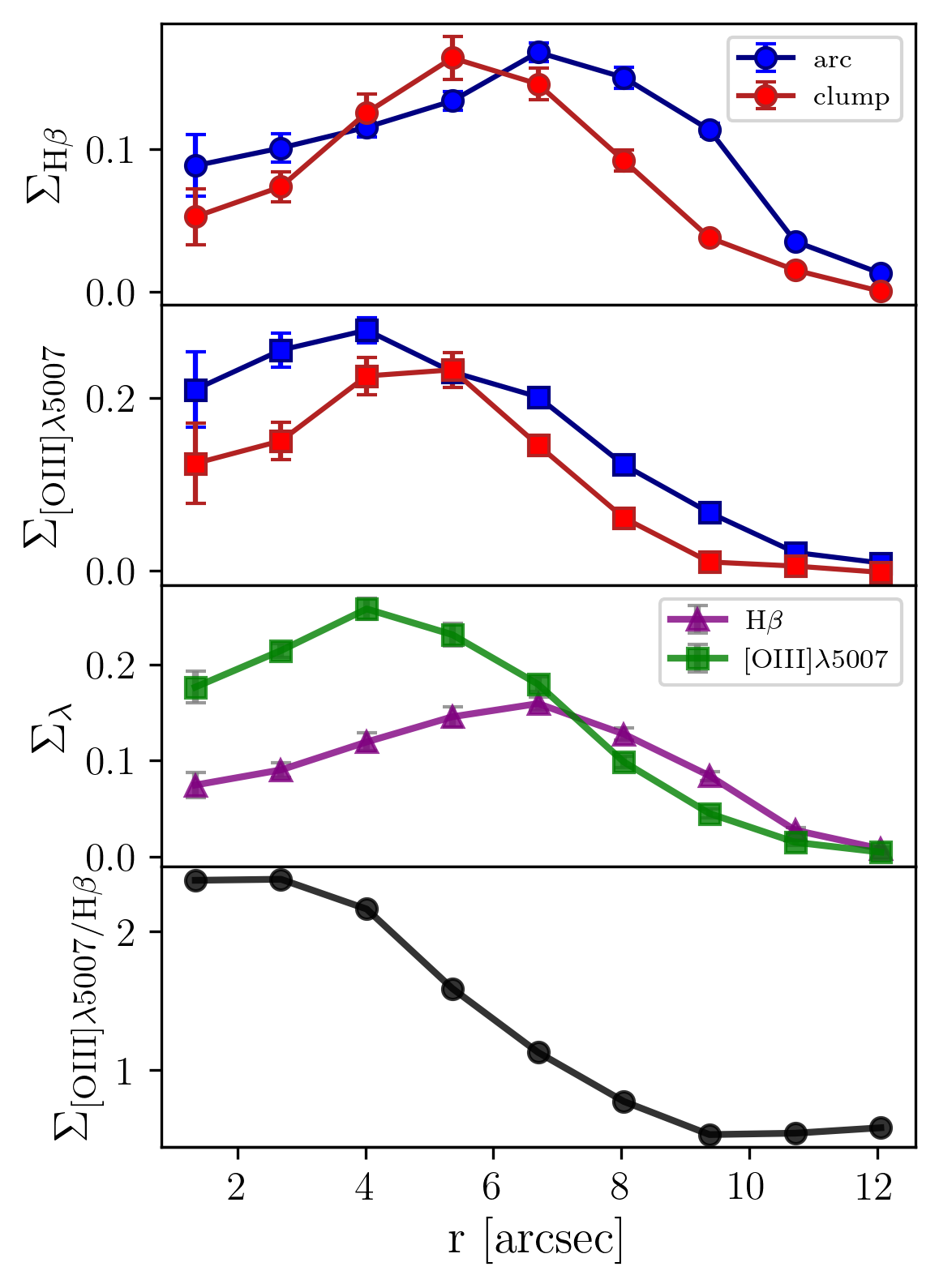}
\caption{Flux density profiles for the south-west clump and the extended arc, with red and blue colours, respectively. The top panel shows the radial H$\beta$ flux density profiles for both structures. The upper-middle panel shows the radial \oiiistrong flux density profile for both structures. The bottom-middle panel shows the H$\beta$ and \oiiistrong flux density profile, with purple and green colours, respectively, for the complete nebula integrated over the entire angular range. The bottom panel shows the \hb/\oiiistrong ratio with black colour for the complete nebula integrated over the entire angular range. }
\label{fig3}
\end{figure}

Figure \ref{fig2} shows the \hb (left panel) and $[\ion{O}{iii}]\lambda5007$ (middle panel) emission line maps for the Leo A \ion{H}{ii} region\textcolor{black}. Line fluxes were measured in each fibre by integrating the Gaussian fits. Concentric circles with radii increasing in steps of 2 px ($1.34''$) up to 20 px ($13.4''$) are plotted for posterior analysis of radial flux profiles. To smooth out spatial variations and artefacts, the maps were smoothed using a bilinear interpolation. The centre of the nebula was defined as the position of the C2 star cluster candidate \citep{stonkute19}.\\
\\
The \hb map (left panel) follows the H$\alpha$ distribution observed with Subaru Suprime-Cam photometry \citet{stonkute14, stonkute19}, tracing a similar morphology as the H$\beta$ emission (red curve in right panel of Figure \ref{fig2}, see also right panel of Figure 3 in \citealt{stonkute19}), suggesting a uniform distribution. This is consistent with the \hb/\hy ratio map, which remains constant across the nebula.

The morphology revealed was decomposed into an area of high emission towards the south-west, and an extended arc with enhanced flux in the north-east. \\
\\
The \oiiistrong map (middle panel) shows a distribution similar to that of ionised hydrogen, with significant emission in the centre. A prominent clump of ionisation is again observed in the south-west at the same location as that detected in the hydrogen map. This clump also exhibits a low H$\beta$/H$\gamma$ ratio (right panel), suggesting a low dust reddening in this region. 
\\
\\
Radial flux-density profiles were constructed to characterise the structures identified in the emission line maps, i.e., the south-western clump and the extended arc. The profiles are defined as $\Sigma_{\lambda} = \sum F_{\lambda}/A_{ring}$, where $\sum F_{\lambda}$ is the integrated flux within a ring of inner and outer radii $r_{in}$ and $r_{out}$, and angular range $\phi_{0}$ and $\phi_{f}$ (counterclockwise in the x-axis). The ring area is  
\begin{equation}
A_{ring} = \int_{\phi_{0}}^{\phi_{f}}\int_{r_{in}}^{r_{out}}rdrd\phi =  \frac{1}{2} (\phi_{f} - \phi_{0})(r_{out}^{2}-r_{in}^{2}) 
\end{equation}
For the arc component, we selected fibres with $40\degree<\phi<290\degree$, while the south-west clump corresponds to $290\degree\leq<\phi<360\degree$ and  $0\degree\leq<\phi\leq40\degree$. The angular selections are presented with the black dotted lines in the left and middle panels of Figure \ref{fig2}. The resulting flux-density profiles, expressed in units of $\mathrm{10^{-16}\  ergs\ s^{-1}\ cm^{-2}\ \AA^{-1}\ \mathrm{arcsec}^{-2}}$, are shown in Figure \ref{fig3}.\\
\\
The upper panel of Figure \ref{fig3} presents the \hb profiles of the arc (blue) and the clump (red). Both components have comparable intensities, with the former peaking at $\sim7''$ and the latter at $\sim5''$. The \oiiistrong profiles (upper middle panel) show similar behaviour, where the arc peaks at $\sim4''$ and the clump peaks at $\sim5''$.\\
\\
The bottom-middle panel compares the \hb and the \oiiistrong profiles for the entire \hii region, i.e., integrating on the complete angular range (from $0\degree$ to $360\degree$). The \oiiistrong emission is more centrally concentrated, while the \hb dominates at larger radii. This behaviour is reinforced by the \hb/\oiiistrong profile (bottom panel), which shows that \oiiistrong prevails over \hb within $\sim7''$. Those features are a classical signature of ionized stratification: due to ionisation feedback (UV radiation from massive stars, stellar winds and SNe explosions; \citealt{osterbrock06}) in a region containing young stellar objects, the surrounding ISM store highly ionized species (such as \opp), whereas the low ionisation species prevail in the outskirts (such as \hp, \op, \np, and \sulfurp), because of the radiation decreases with distance.\\
\\
These findings are similar to those observed in the SagDIG \hii region \citep{andrade25}. A direct confirmation of stratification requires comparing \op and \opp distributions using \oiiistrong and \oii emission lines, respectively. However, the VIMOS data do not cover wavelengths bluer than $4150\ \AA$, and two-dimensional calibrated long-slit observations are not available for this nebula. Fortunately, indirect evidence is provided by the C2 star cluster candidate (age $\sim20$ Myr and $M_{*}\geq 150 \mathrm{M_{\odot}}$) detected by \citet{stonkute19}, based on HST/ACS and the Subaru Suprime-Cam photometry (see their Figure 3). This young stellar object is located at the centre of the nebula. \\
\\
The ionised stratification in \hii regions should have stellar systems responsible for bringing radiation to the ISM, so the location of C2 with the \hp and \opp distributions suggests that stellar feedback from this cluster is the primary source of ionisation and mechanical energy input, thus producing the stratified distribution. Moreover, if this is the case, the structure of the nebula may represent a classical bubble-like ionised-bound \hii region \citep{pagel97, osterbrock06}.

% ----------------------------------------------------------------------
% ----------------------------------------------------------------------
% ----------------------------------------------------------------------
% ----------------------------------------------------------------------
% ----------------------------------------------------------------------
% ----------------------------------------------------------------------
% ----------------------------------------------------------------------
% ----------------------------------------------------------------------
% ----------------------------------------------------------------------
% ----------------------------------------------------------------------
% ----------------------------------------------------------------------
% ----------------------------------------------------------------------
% ----------------------------------------------------------------------
% ----------------------------------------------------------------------
% ----------------------------------------------------------------------
\subsection{Electron temperature estimates}
\label{temperatures}
We estimated the gas-phase metallicity using the so-called "direct method" \citep{peimbert67, aller84}, which relies on electron temperature-sensitive lines together with the respective line emissivities. \\
\\
Electron temperatures were derived via nebular-to-auroral transitions of \opp, i.e., $[\ion{O}{iii}]\lambda\lambda4959+5007/4363$. Because the \oiiiauroral line is typically 2-3 orders of magnitude fainter than \hb \citep{maiolinomanucci19}, it is often affected by spectral noise. The abundance determination also requires estimating the electron density, $n_{e}$, generally obtained from $[\ion{S}{ii}]\lambda\lambda6717,6731$. However, since our spectra do not extend beyond $6200\AA$, we adopted $n_{e}=100$ cm$^{-3}$ according to the low density limit \citep{hummer87}.\\
\\
We computed $T_{e}[\ion{O}{iii}]$ using the nebular-to-auroral ratio with the \textit{getTemDen} module of Pyneb \citep{luridiana15}. Then, the corresponding $T_{e}[\ion{O}{ii}]$ was obtained from the relation $T_{e}[\ion{O}{ii}] = 0.7\times T_{e}[\ion{O}{iii}]+3000$ given by \citet{campbell86}.\\
\\
For the integrated spectrum of the Leo A \hii region, we derived $T_{e}[\ion{O}{iii}] = 22055\pm2052$ K. Using the mock slit fluxes, consistent results are obtained: $T_{e}[\ion{O}{iii}] = 22693\pm426$ K (VZ06 mock slit) and
$T_{e}[\ion{O}{iii}] = 22332\pm2047$ K (R18 mock slit). Those values agree within the uncertainties, suggesting no significant temperature fluctuations across the nebula. 

\subsection{$T_{e}-$based oxygen abundances}
\label{metallicities}
The total oxygen abundance is defined as $12+\log(\mathrm{O/H})$, where $\mathrm{O/H = (O^{+}/H^{+} + O^{++}/H^{+})}$. Because the VIMOS-IFU spectral range does not include \oii, we reproduce the \citetalias{vanzee06} and the \citetalias{ruizescobedo18} long-slit observations within our data cube to incorporate their \oii fluxes (see Appendix \ref{appendix:mock_slits}). In addition, we also found discrepancies in the \ophp ionic abundance estimates, which come from \citetalias{vanzee06} and \citetalias{ruizescobedo18} flux measurements, where the spectrum of the latter (their Figure 4) shows that \oii falls in the region with higher noise, likely affecting their flux measurements. This is not observed in the \citetalias{vanzee06} spectrum (their Figure 2), where \oii is clearly detected. For this reason, and to provide an accurate measurement of the total $T_{e}$-based oxygen abundance, we used the \citetalias{vanzee06} \oii flux to derive \ophp and the VZ06 VIMOS mock slit to derive \opphp. The detailed procedure, involving the combination and consistency checks, is described in Appendix \ref{appendix:te}.
\\ 
\\
The $T_{e}$-based metallicity is $12+\log(\mathrm{O/H})=7.29\pm0.06$ dex (VZ06 mock slit). In addition, we also derived metallicities from the $R_{23}$ strong-line index \citep{kk04} returning $12+\log(\mathrm{O/H})=7.44\pm0.07$ dex, in agreement with the empirical ($\sim7.48$ dex) and the semiempirical ($7.44 \pm 0.10$ dex) estimates reported in \citetalias{vanzee06}.
\\
\\
In Section \ref{temperatures}, we showed that the three $T_{e}$ estimates (from both mock slits and the integrated spectrum) does not show significant $T_{e}$ fluctuations. However, they could be present as a thermal gradient (e.g., \citealt{andrade25}), but below the VIMOS-IFU sensitivity. Therefore, we take into account possible temperature fluctuations \citep{kobulnicky96} in the abundance determination, which is a well known systematic bias in when metallicities are derived using the $T_{e}-$sensitive method \citep{peimbert67,peimbert17}.\\
\\
We applied the correction proposed by \citet{cameron23}, which increases the abundance to $12+\log(\mathrm{O/H})\ = 7.46 \pm 0.09$. This correction should be carefully taken, as it was calibrated for higher metallicities and lower $T_{e}$ regimes. Despite this, the correction does not affect our interpretations regardless of whether the corrected or uncorrected abundance is adopted. The procedure of the $T_{e}$ corrections is described in Appendix \ref{appendix:te_corrections}.

\section{Discussion}
\label{discussionssss}
\subsection{Stellar distribution in the the Leo A \ion{H}{ii} region}
\label{hstcomparison}

The stellar populations of Leo A have been extensively studied using HST/ACS and Subaru Suprime-Cam photometry \citep{cole07, stonkute14}. The CMD reveals a well-defined red giant branch (RGB), red clump, and blue plume, together with a small population of asymptotic giant branch (AGB) stars \citep{cole07,stonkute14, lescinskaite21,lenscinskaite22}. This morphology reflects an extreme case of a "late-blooming" SFH, as the case of Aquarius, WLM, and Leo P (\citealt{cole14, mcquinn24a, mcquinn24b}): an initial episode more than 10 Gyr ago was followed by a $\sim2$ Gyr quiescent phase, after which the star formation was reignited and continued from $\sim$8 Gyr ago to the present day, peaking at rates 5-10 times higher than the present. This SFH \citep{cole07} is consistent with the behaviour of "slow dwarfs" in the \citet{gallart15} framework and with the "two-component" classification of \citet{benitez15}.\\ 
\\
We relate this stellar context to the properties of the Leo A \hii region. For this purpose, we used the HST/ACS photometry \footnote{\href{https://mast.stsci.edu/portal/Mashup/Clients/Mast/Portal.html}{Hubble Source Catalog (HSCv3)}}, which provides better spatial resolution compared with the Subaru data. This allows us to directly connect the location of the young stellar population with the ionised gas structure observed in the VIMOS-IFU cube. 

\begin{figure*}
\centering
\includegraphics[width=\hsize]{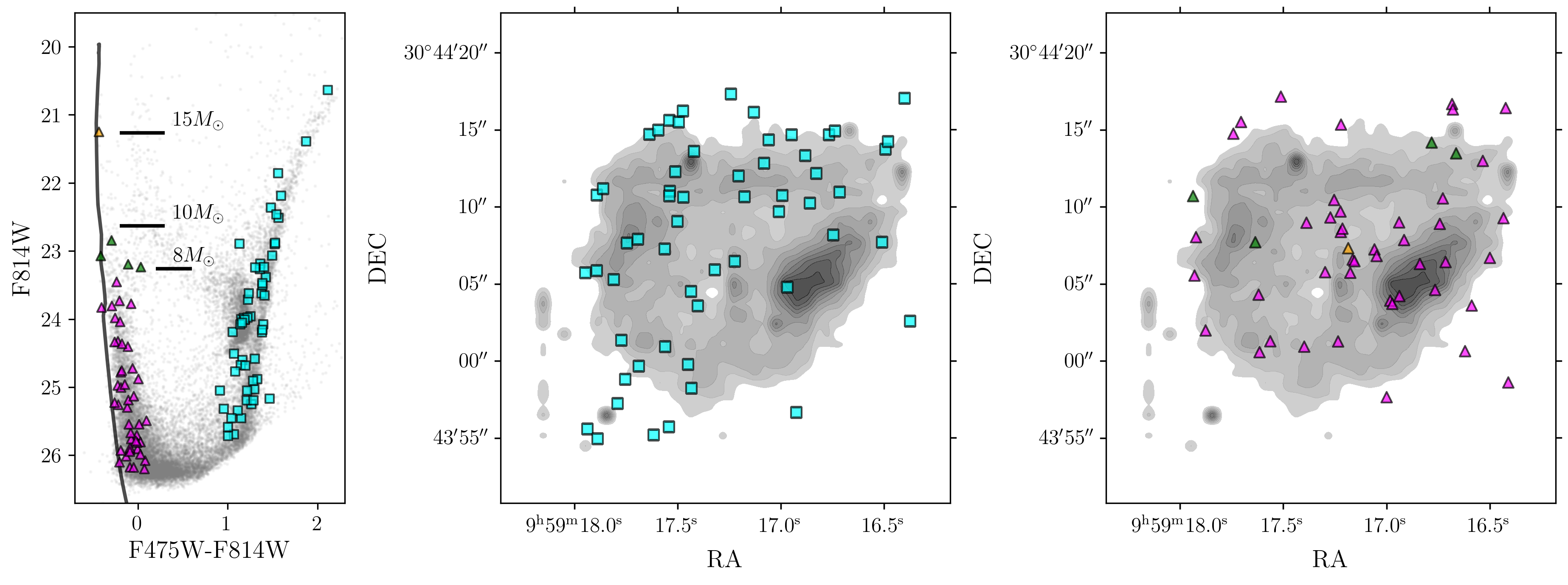}
\caption{Comparison of the Leo A \hii region with H$\beta$ emission (grey contours) with HST/ACS photometry. The left panel shows the location of young MS stars (magenta triangles) and old stars (cyan squares). A 10 Myr PARSEC isochrones is shown as the black curve, in order to get a proxy of the stellar masses of the young stars. The middle panel shows the spatial distribution of old stars in the \hii region, whereas the right panel shows the distribution of young MS stars in the \hii region. The most massive star is shown with the orange triangle, and those with masses $8< M_{\odot}<10$ are shown with green triangles}.
\label{fig7}
\end{figure*}

To analyse the spatial distribution of stellar populations within the Leo A \hii region, we applied a similar approach as in \citet{andrade25}. PARSEC v1.2S theoretical isochrones were fitted in the HST/ACS photometry, and colour cuts were used to classify stars into young MS and older populations (RGB and red clump stars), as described in Appendix \ref{appendix:CMD}. The resulting classification is presented in Figure \ref{fig7}: young MS stars are shown with magenta triangles, old stars as cyan squares, and the full HST catalogue as grey points for reference.\\
\\
The middle and right panels of Figure \ref{fig7} present the spatial distribution of these stellar populations relative to the \hb emission (grey contours). The old stars are distributed uniformly across the \hii region, but avoiding the south-west ionisation clump. In contrast, young MS stars are found all over the nebula, but with some preference towards the centre and the ionisation clump.\\
\\
To get a proxy of stellar masses, we compared the young MS population with a 10 Myr isochrone (grey curve in the left panel of Figure \ref{fig7}). Four stars consistent with masses of $\sim8-10M_{\odot}$ (green triangles) are located along the arc structure, while the most luminous star (yellow triangle), with an estimated mass of $\sim15M_{\odot}$ is found at the nebular centre.

\subsection{Sources of ionisation in the Leo A \hii region}
\label{ionisation}

\begin{figure*}
\centering
\includegraphics[width=\hsize]{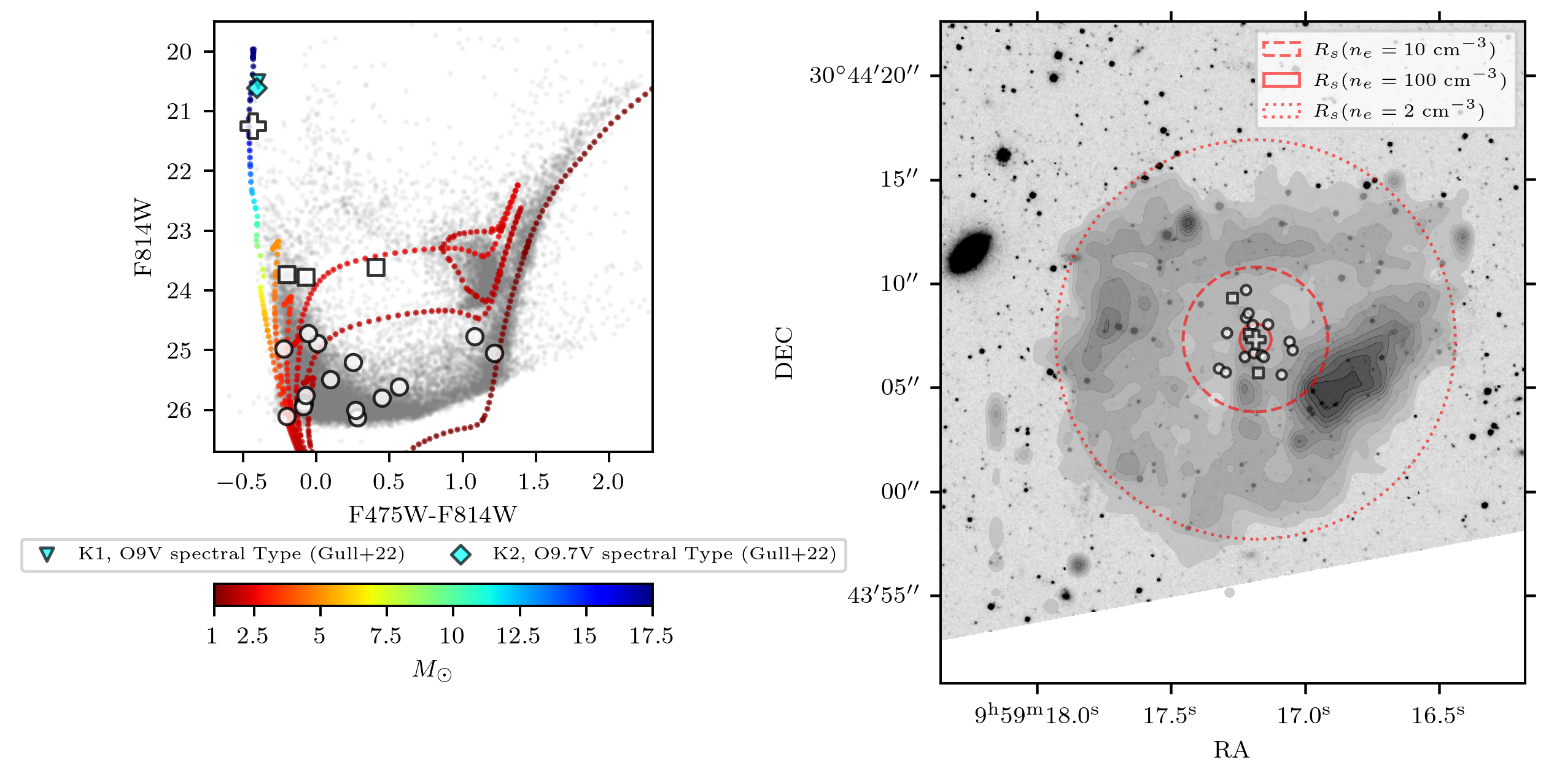}
\caption{Location of stars belonging to the C2 star cluster ($<2.5''$). Left panel: location of stars in the CMD. The white cross corresponds to the brightest star. The white squares are those stars with $23<$F814W$<24$. Those stars with F814W$>24$ are shown as white dots. Theoretical PARSEC isochrones of  10 Myr, 100 Myr, 250 Myr, 500 Myr,1 Gyr, and 5 Gyr are shown as colour-coded curves according to their stellar masses given by the theoretical isochrones. K1 and K2 O-type stars from \citet{gull22} are shown as the cyan triangle and the cyan diamond, respectively. Those massive stars are capable of ionising individually two Leo A \hii regions. Right panel: location of the star of the C2 star cluster in the Leo A \hii region (grey contours), superimposed with the HST/ACS F814W frame.}
\label{fig8}
\end{figure*}

Massive stars are the primary ionising sources of \hii regions, through UV radiation, stellar winds, or via SN explosions \citep{osterbrock06}.\\
\\
\citet{gull22} reported tentative evidence that two out of four \hii regions in Leo A (in the eastern sector) are powered by their central stars, labelled K1 and K2 with spectral types O9V and O9.7V, respectively. Their effective temperatures ($T_{\mathrm{eff}} \sim 30900$ and $\sim31600$ K) and ages ($\sim$10 and $\sim$8 Myr) produce photionisation production rates consistent with the approximations presented in nebular analyses \citepalias{vanzee06, ruizescobedo18}.\\
\\
The \hii region analysed in this work was not included in \citet{gull22}. Therefore, we investigate whether a single star of a clump of stars is ionising the \hii region.\\
\\
At the nebular centre of this \hii region is found the C2 star cluster candidate, identified by \citet{stonkute19}, with age $\sim20$ Myr and $M\geq150 M_{\odot}$. We consider stars within $2.5''$ (white symbols in the right panel of Figure \ref{fig8}) of the cluster centre as members of C2 \citep{stonkute19}. The CMD analysis (left panel of Figure \ref{fig8}) shows that the brightest star (white cross) is consistent with a mass of $\sim$15 $M_{\odot}$ and an age of $\sim$10 Myr, while most of the surrounding stars (white squares and dots) have masses below $5\ M_{\odot}$, suggested by the colour-code of the PARSEC isochrones. The isochrones predict $T_{\mathrm{eff}}\sim33200$ K, comparable to K1 and K2 values reported by \citet{gull22}, as also indicated by their location in the CMD (cyan triangle and diamond, respectively). This indicate that the most luminous star may be the dominating source of ionisation for the nebula.

To test this hypothesis, we estimated the stellar photoionisation produced rate, $Q$, using the PARSEC v2.0 evolutionary tracks \citep{nguyen22, costa25}. We took the mean $Q$ value for stellar tracks between $M\in[14,16]$ $M_{\odot}$, $Z\in[0.006,0.008]$, and $T_{\mathrm{eff}}\in[30000,35000]$ K, which resutls in $Q=10^{46.9\pm0.17}$ photons s$^{-1}$. Uncertainties were computed by taking the standard deviation of the Q values with the filters applied.\\
\\
Then, an independent $Q$ estimate was derived from nebular parameters. Assuming Case B recombination \citep{hummer87, storey95}, the relation between $Q$ and H$\alpha$ luminosity is $Q=7.315\times 10^{11}L(\mathrm{H}\alpha)$ \citep{kennicutt98, choi20}. To estimate $L(H\alpha)$, the theoretical relation $I(\mathrm{H\alpha})=2.86\ I(\mathrm{H\beta})$ was used \citep{hummer87}. We derive $Q=10^{47.17\pm0.08}$ photons s$^{-1}$. This value agrees within uncertainties with both the stellar estimate and the H$\alpha$-based determination of \citetalias{ruizescobedo18}, $Q=10^{47.16\pm0.01}$ photons s$^{-1}$.\\
\\
This consistency between the stellar and nebular $Q$ estimates supports the conclusion that the central star of C2 is the main ionising source of this \hii region. \\
\\
The Str$\ddot{\mathrm{o}}$mgren radius was estimated as follows: 

\begin{equation}
\label{str_radius}
    R_{s} = \left( \frac{3Q}{4\pi n_{e}^{2}\alpha_{B}(T)} \right)^{\frac{1}{3}}
\end{equation}
where $n_{e}$ is the electron density and $\alpha_{B}(T)$ is the recombination coefficient at a given temperature. We adopted the empirical description of $\alpha_{B}(T)$ from \citet{lequeux05} using our measured $T_{e}$, and under $n_{e}=100$ cm$^{-3}$. This results in $R_{s}=3.2\pm0.2$ pc (red circle in Figure \ref{fig8}), consistent with the 2.7 pc derived by \citetalias{ruizescobedo18} from photoionisation models.

However, the observed size of the nebula is $\simeq94$ pc\footnote{Based on measuring the pixel length along the South-East to North-West diagonal (36 px) in the H$\beta$ emission line map, generated by selecting all fibers where H$\beta$ and \oiiistrong have a S/N $>3$. We adopted the Leo A distance of 800 kpc \citep{dolphin02}.} ($R\simeq 47$ pc), compatible with the physical scale of the Leo A galaxy presented in \citet{lenscinskaite22}. Thus, the brightest star seems not to be enough to sustain ionisation of the complete \hii region.\\
\\
\citetalias{ruizescobedo18} gives and estimate of $n_{e}$ as $131\pm121$ cm$^{-3}$, using the $[\ion{S}{ii}]\lambda\lambda$ doublet, which is in line with the size-density relation of extragalactic \hii regions of \citet{hunt09}. Hence, we relaxed the $n_{e}$ assumption. In this case, we derived $R_{s}=15\pm1$ pc (red dashed circle in Figure \ref{fig8}) under $n_{e}=10$ cm$^{-3}$. This implies that the central star can only ionise $\sim32\%$ of the nebula. On the other hand we found that at $n_{e}=2$ cm$^{-3}$, $R_{s}=41 \pm 3$ pc, consistent with the radius of the \hii region (red dotted circle in Figure \ref{fig8}).\\
\\
The agreement between $Q$ derived from stellar tracks and nebular parameters suggest that the star is, indeed, the dominant ionising source, similar to the K1 and K2 stars powering other \hii regions of Leo A \citep{gull22}, and analogous to the case of Leo P, where a single massive star sustain with ionisation its only \hii region \citep{telford22}. However, depending of the adopted $n_{e}$ the interpretations can change: if $n_{e}\geq10$ cm$^{-3}$, the star is not able to ionise completely the \hii region, and additional ionising sources are required to explain the full size of the \hii region, supported by the presence of the C2 star cluster. In this case the ionisation photon budget should be $Q\geq 10^{48.7}$ photons s$^{-1}$ (Eq. \ref{str_radius}). On the contrary, if $n_{e}<10$ cm$^{-3}$, the brightest star is able to ionise completely the \hii region, which is supported by the fact that both $Q$ estimates (the extracted by the stellar tracks, and the obtained with nebular parameters) agrees.\\
\\
It is necessary to stress that these are rough approximations, since we assumed radial symmetry, uniform $T_{e}$, and uniform $n_{e}$ distributions, which is not the case since emission line maps (Figure \ref{fig2}) clearly show significant spatial variations.

\subsection{The Leo A \ion{H}{ii} region in the mass-metallicity plane}

\begin{figure}
\includegraphics[width=\hsize]{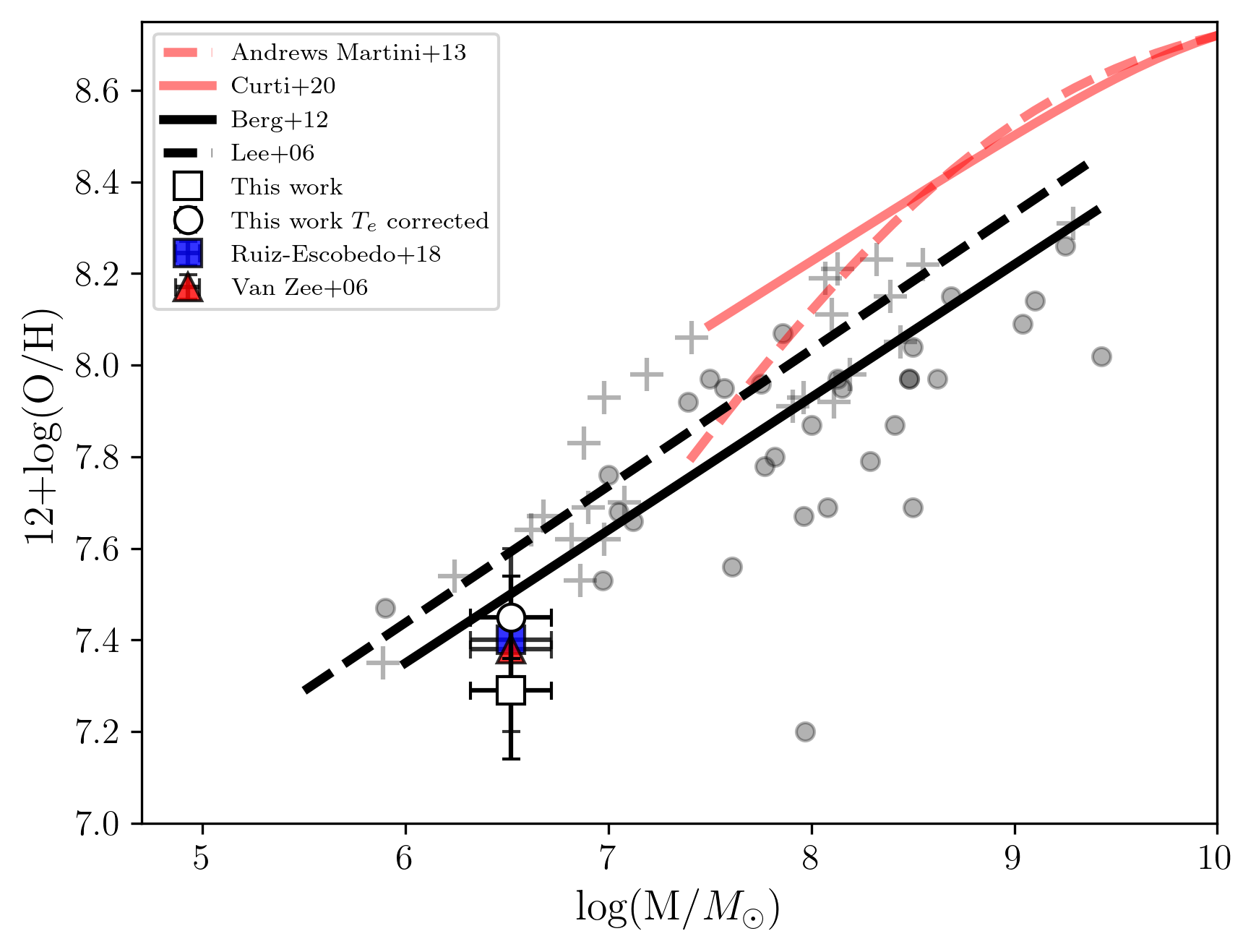}
\caption{Leo A in the mass-metallicity plane, with white square and white dot representing the uncorrected and corrected metallicities for $T_{e}$ fluctuations, respectively. The \citetalias{vanzee06} ($R_{23}$) and \citetalias{ruizescobedo18} ($T_{e}-$based) measures were included as the red triangle and blue square, respectively, for comparison. Black dashed and black solid lines are the \citet{lee06} and the \citet{berg12} low-mass end of the MZR, respectively. The grey crosses and the grey dots are their respective samples. The solid red curve and the dashed red curve are the MZR of the local universe of \citet{andrewsmartini13} and \citet{curti20}, respectively.}
\label{fig9}
\end{figure}

The spatial analysis of the Leo A \hii region with VIMOS-IFU, combined with HST/ACS photometry, suggests that the structure is strongly influenced by stellar feedback from the C2 cluster. The impact of stellar feedback can also be constrained by the chemical properties of the nebula.\\
\\
\citetalias{vanzee06} reported that the four \hii regions of Leo A exhibit similar gas-phase metallicities ($12+\log\mathrm{(O/H)}\sim7.40$) based on the empirical $R_{23}$ calibrator. We therefore consider our $T_{e}$-based total oxygen abundance for the nebula studied as representative of the galaxy as a whole. Adopting $3.3\pm 0.7 \times 10^{6} \mathrm{M_{\odot}}$ from \citet{kirby17}, we consider both uncorrected and corrected metallicities for $T_{e}$ fluctations to place Leo A in the mass-metallicity plane, because most of the works that apply this method do not take into account this bias.\\
\\
Figure \ref{fig9} shows the position of Leo A in the mass-metallicity plane. The uncorrected and corrected values for $T_{e}$ fluctuations are shown as the white square and dot, respectively. On the other hand, the \citetalias{vanzee06} $R_{23}$-based abundance and the \citetalias{ruizescobedo18} $T_{e}$-based abundance are shown with the red triangle and blue square, respectively. These are compared with the $T_{e}$-based mass-metallicity relation (MZR) of the local universe of \citet{andrewsmartini13} and \citet{curti20}, as red solid and red dashed curves, respectively. We also compared with the low-mass end of the MZR of \citet{lee06} and \citet{berg12}, as black dashed and solid lines, respectively.

Our estimates place Leo A in agreement with the low-mass end of the MZR, within the 0.15 dex scatter of \citet{lee06}. 

The nature of the low-mass end of the MZR is still matter of debate. Observations show that the scatter in the low-mass regime \citep{zahid12} reflects the interplay of gas accretion, outflows, star formation, and environmental dependency, with the relative importance of those mechanisms varying on a galaxy-by-galaxy basis (e.g., \citealt{dalcaton04, petropoulou12, chisholm18, duartepuertas22}). In theoretical frameworks, the picture is also diffuse, since the low-mass end of the MZR can be reproduced, for example, by (i) SNe driven winds (e.g., \citealt{finlatordave08, dave12, guo16}), (ii) gas regulation by material flows plus star formation \citep{lilly13}, (iii) gas-infall dominated galaxies where star formation rate obeys the Kennicutt-Schimdt law subject to a gas density threshold \citep{tassis08}, or (iv) variations in the initial-mass function (IMF, \citealt{koppen07}).

\subsection{Chemical evolution models of Leo A}
\label{chem_models_leoa}
\begin{figure*}
\includegraphics[width=\hsize]{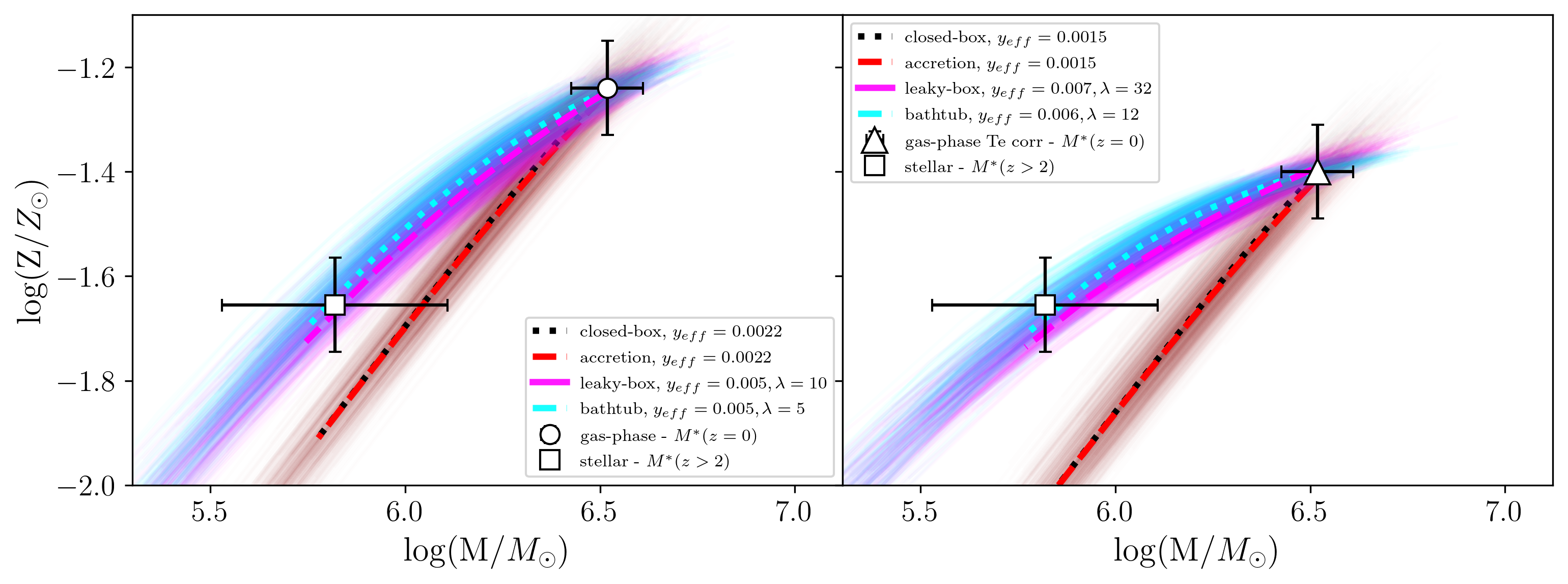}
\caption{Evolution of Leo A in the last 10 Gyrs in the mass-metallicity plane. Left panel: The white square represents Leo A 10 Gyr ago, whereas the white dot represents Leo A in the present day. The closed-box and accretion model median tracks that approach our estimates are shown as the black dotted and red dashed lines, respectively. The Leaky-box and the gas regulator (bathtub) model median tracks that reproduce our estimates are shown with the magenta solid and the cyan dashed curves, respectively. The closed-box, accretion, leaky-box and gas regulator tracks generated by the Monte Carlo simulations are shown with the semi-transparent black, red, magenta, and cyan curves, respectively. Right panel: same as the left panel, but using the $T_{e}$ corrected gas-phase metallicity as the chemical status of Leo A at the present day, with the white triangle.}
\label{fig10}
\end{figure*}

\citet[hereafter K17]{kirby17} using chemical evolution models applied to the stellar metallicity distribution of Leo A, suggested that the galaxy was either pre-enriched or acquired external gas during its SFH \citepalias[Figure 9 and 10]{kirby17}. In addition, our results suggest that the ionised gas seems to be driven by stellar feedback. To test whether both perspectives are part of the same picture, we explore simple chemical evolution models with a similar approach as that presented in \citet{barrera18}, and \citet{olvera24}.\\
\\
Old stars store in their atmospheres the chemical composition of their natal clouds. We therefore compared the present-day gas-phase metallicity of Leo A\footnote{$\mathrm{\log(Z/Z_{\odot})=12+\log(O/H)-8.69 = -1.39 \pm 0.06}$}$^{,}$\footnote{$\mathrm{\log(Z/Z_{\odot})=12+\log(O/H)_{Te-corrected}-8.69 = -1.23 \pm 0.09}$} with the mean stellar metallicity\footnote{$\mathrm{\log(Z/Z_{\odot})=[Fe/H]+0.75[\alpha/Fe]}=-1.65\pm0.09$ \citep{kirby17}} of its old stellar population of  $\left< [\mathrm{Fe/H}] \right> = -1.67^{+0.09}_{-0.08}$ \citep{kirby17}.\\
\\
To properly apply the models, we also consider their stellar mass. Leo A has a present-day stellar mass of $3.3\pm0.7\times10^{6}\ M{\odot}$ \citep{kirby17}, of which 
$0.66^{+0.91}_{-0.44}\times10^{6}\ M_{\odot}$ was created 10 Gyr ago ($z>2$), as estimated by \citet{bermejo18} by integrating the Leo A SFH. \\
\\
We assume that the mean stellar metallicity measured by \citetalias{kirby17} reflects the chemical state of Leo A 10 Gyr ago. This assumption is supported by observational evidence \citepalias[Figure 12]{kirby17} and by models of the age-metallicity relation (AMR), which remains flat since the beginning of star formation up to $\sim5$ Gyr ago \citep{hidalgo17, ruizlara18}.\\
\\
Analytic chemical evolution models describe the gas-phase metallicity as $Z_{g}=Z(\mu,y)$, a function of the gas fraction $\mu=M_{\mathrm{gas}}/(M_{*}+M_{\mathrm{gas}})$ and the yield $y$, defined as the mass fraction of metals produced by a stellar generation, relative to the mass fraction locked up in stellar remnants and long lived stars, for a given IMF\footnote{The yield, $y$, is described as $\int_{1}^{\infty}m p_{m}\phi(m)dm/(1-R)$. Where $p_{m}$ is the yield of metals per stellar mass, $\phi(m)$ is the IMF, and $R$ is the returned fraction, as the fraction of mass ejected into the ISM by a entire stellar generation \citep{tinsley80}.} \citep{matteucci21}. More realistic scenarios incorporate metal-poor inflows and metal-rich outflows.\\
\\
We follow the evolution of the gas fraction with time. At the present day, Leo A has a gas mass of $6.9\pm0.7\times10^{6}\ M_{\odot}$ \citep{hunter12} and a stellar mass of $3.3\pm0.7\times10^{6}\ M_{\odot}$ \citepalias{kirby17}, resulting in $\mu(t=0)=0.68\pm0.05$.
On the other hand, 10 Gyr ago, adopting the same gas mass and the stellar mass 10 Gyr ago of \citet{bermejo18}, the gas fraction was $\mu(t=10\ \mathrm{Gyr})=0.92\pm0.06$. In both cases, the same gas mass was considered for consistency with the \citetalias{kirby17} results, i.e., the stellar populations are consistent with a pre-enriched closed-box or an accretion scenario for the evolution of Leo A.

We parametrise the evolution of the gas fraction with time with a linear interpolation, $\mu(t)=\alpha t+\mu(t=0)$, with $t\in[0,10]$ Gyr, and $\alpha$ as a free parameter. Therefore, the changes in the gas fraction with time track the build-up of stellar mass across cosmic time. \\
\\
We tested four models of chemical evolution. The simplest is the so-called closed box model \citep{searle72}, where a galaxy is treated as a system with no material flows into or out of the galaxy. The chemical enrichment is driven only by star formation. In this framework, part of the gas expelled by SNe explosions is used to form new generations of stars, and the remaining gas is used to enrich the ISM. The gas metallicity evolves as:

\begin{equation}
\label{eq:cb}
    Z_{g} = y \ln(\mu^{-1})
\end{equation}
where $y$ is the yield, and $\mu$ is the gas fraction. Under this approximation, the metallicity asymptotically approaches to $Z_{g}= y$ at the end of the SFH of a system.
\\ 
\\
The second is the accretion model \citep{larson72, tinsley80} which includes continuous infall of metal-poor gas. The metallicity follows:

\begin{equation}
\label{eq:acc}
    Z_{g} = y_{\mathrm{eff}} \left[ 1- e^{\left(1-\mu^{-1}\right)} \right]
\end{equation}
where \yeff is known as the effective yield, which is lower than the true yield of the closed-box model. In this scenario, and the next model introduced, \yeff represents the highest degree of chemical enrichment for a given IMF \citep{matteucci21}.\\
\\
The third case is the leaky-box model \citep{matteucci83}, for a galaxy suffering mass loss through outflows. The metallicity is given by
\begin{equation}
\label{eq:lb}
    Z_{g} = \frac{y_{\mathrm{eff}}}{1+\lambda} \ \ln{\left[ (1+\lambda)\mu^{-1} - \lambda \right]}
\end{equation}
where $\lambda$ is the mass loading factor, and quantifies the amount of gas expelled from the galaxy relative to star formation.\\
\\
Finally, we considered the gas-regulator or "bathtub" model of \citet{lilly13}. Here, the global SFR across cosmic time is regulated by variations of the gas reservoir subject to inflows, outflows, gas used to form long-lived stars, and instantaneous recycling to enrich the ISM. The metallicity is expressed as: 
\begin{equation}
\label{eq:bt}
    Z_{g} = \frac{y_{\mathrm{eff}}}{1+r_{gas}+(1-R)^{-1}\left(\lambda + \epsilon^{-1} \frac{d\ln{(r_{gas})}}{dt} \right)}
\end{equation}
where $r_{gas}=\mu/(1-\mu)$ is the gas-to-stellar mass ratio, $1-R$ is the amount of gas used to create long-lived stars, and $\epsilon$ is the star formation efficiency. Following \citet{lilly13} and \citet{barrera18} we used $R=0.4$, and $ \epsilon^{-1} \frac{d\ln{(r_{gas})}}{dt}\sim -0.25$.\\
\\
To apply these solutions, we evolved the gas fraction between 10 Gyr ago and the present day using the linear interpolation described above. For each model, we ran Monte Carlo simulations, sampling $\mu(t=0)$ and $\mu(t=10)$ from Gaussian distributions centred on their estimated value with $\sigma$ settled by their uncertainties. For the closed-box and accretion models, we ran 1000 simulations per \yeff from 0.001 to 0.02 (steps of 0.001). For the leaky-box and the gas-regulator models, we ran 1000 simulations over the same \yeff combined with $\lambda$ values from 0 to 35 (steps of 1).\\
\\
Figure \ref{fig10} compares the chemical state of Leo A at the present day and 10 Gyr ago. The present-day gas-phase abundances, with $T_{e}$ correction and without correction, are shown as the white dot (left panel) and white triangle (right panel), together with the stellar metallicity (white square in both panels).\\
\\
Using the $T_{e}$ corrected gas-phase metallicity (left panel), the closed-box (black dotted) and accretion models (red dashed) require \yeff$=0.0022 \pm 0.0005$ to match the present-day abundance. However, these scenarios fail to reproduce the full chemical evolution of Leo A, as their mean evolutionary tracks are not consistent with the stellar and gas-phase metallicity constraints. On the other hand, the leaky-box model (magenta dashed) with \yeff$=0.005\pm0.001$ and $\lambda=10$, and the gas-regulator model (cyan dotted) with \yeff$=0.005\pm0.001$ and $\lambda=5$, produce tracks that successfully connect the chemical status of Leo A 10 Gyr ago and at the present day. The Monte Carlo simulations are shown with their respective colours as semi-transparent curves.\\
\\
Using the non $T_{e}$-corrected gas-phase abundance (red panel), the results are similar: closed-box and accretion scenarios fail to reproduce the Leo A chemical evolution, while the leaky-box ($y_{\mathrm{eff}}=0.007\pm0.002$, $\lambda\sim32$) and accretion models ($y_{\mathrm{eff}}=0.006\pm0.002$, $\lambda\sim12$) successfully reproduce the estimated evolution. In both $T_{e}$-corrected and uncorrected cases, the key parameter is the inclusion of mass loss.\\
\\
In the mass-metallicity plane, closed-box and accretion models behave similarly. The same is observed with the leaky-box and gas-regulator models. Those features comes from the functional form in the solutions of the four models. The detailed procedure exploring these features is presented in Appendix \ref{appendix:chem_models}.\\
\\
Previous works, as \citet{garnet02} suggested \yeff$\simeq0.002$ in the closed-box framework, while \citetalias{kirby17} derived \yeff$\simeq 0.005$ under a pre-enriched or accretion scenario. Our calculations closely agree with the latter, and are consistent with \yeff of dwarf galaxies in the Local Universe with comparable masses \citep{garnet02,tortora22}.\\
\\
Leaky-box and gas-regulator models indicate that outflows must be considered in the evolution of Leo A. The mass loading factor that reproduces the chemical evolution of Leo A differs between them. This difference lies in their main assumptions: the leaky-box describes a closed-box with outflows, while the gas-regulator incorporates the balance of inflows, outflows, and star formation, regulating the gas reservoir. \\
\\
Based on the fact that the Leo A evolution is being described by a leaky-box model, the simplest interpretation is that Leo A is driven by galaxy-scale outflows. However, the evolution of gas fractions was constructed under constant accretion, as \citetalias{kirby17} suggests. Therefore, our leaky-box model is an accretion model with material loss driven by stellar feedback. In other words, our leaky-box model is a rough approximation of the gas-regulator model, explaining why both reproduce the observed evolution with similar tracks but different values of $\lambda$. \\
\\
Our interpretation is therefore that Leo A was either pre-enriched or acquired significant gas during its early star formation, allowing for the development of its stellar metallicity distribution. At the present day, however, stellar feedback and gas loss play a significant role. The present-day evolution of Leo A appears to be governed by a balance between gas flows and star formation, i.e., a gas equilibrium framework (e.g., \citealt{finlatordave08, tortora22}), under a roughly constant \yeff from 10 Gyr to the present day. In this context, the conclusions of \citetalias{kirby17} and our results are complementary, providing together a coherent picture of the chemical evolution of Leo A across cosmic time. 

\section{Summary and conclusions}
We analysed intermediate-resolution optical VIMOS-IFU/VLT archival data of one of the four \hii regions in Leo A. We explored the nebular morphology through emission line maps. We derived $T_{e}$-based total oxygen abundances, and we used HST photometry to link young stellar populations with the gaseous structure, and tested chemical evolution models of Leo A over the past 10 Gyr.\\
\\
Our main conclusions can be summarized as follows:

\begin{itemize}
    \item The \hb map shows weak central emission, whereas \oiiistrong remains strong at the centre, indicating ionised stratification, with \opp more centrally concentrated than \hp.

    \item The young MS stars found in the nebular centre of the \hii region correspond to the C2 star cluster identified by \citet{stonkute19}. 

    \item The brightest MS star, located in the nebular centre, seems to be an O-type star with $\sim15\mathrm{M_{\odot}}$ and $T_{\mathrm{eff}}\sim33200$ K.
    
    \item By extracting mock slits following \citetalias{vanzee06} and \citetalias{ruizescobedo18}, we measured $T_{e}=22693\pm436$ K for the \hii region. The $T_{e}$-based derived metallicity is \mets$= 7.29\pm 0.06$ dex, while the $R_{23}$-based metallicity derived is \mets$= 7.45\pm 0.07$, in agreement with the empirical and semi-empirical values reported by \citetalias{vanzee06}.

    \item Applying the \citet{cameron23} calibration to correct for $T_{e}$ fluctuations, we derived $T_{e}= 16431 \pm 1203$ and a metallicity of \mets$=7.45\pm 0.09$, being $\sim0.16$ dex higher than the uncorrected value.

    \item The leaky-box and gas-regulator models reproduce the stellar mass growth ($\sim2.4\times10^{6},M_{\odot}$) and chemical enrichment ($\sim0.4$ dex) over the last 10 Gyr with $y_{\mathrm{eff}}\sim0.005\pm0.001$, consistent with \citetalias{kirby17}, a value slightly larger than the closed-box approximation presented in previous works \citep{garnet02} but in line with estimates of other dwarf galaxies in the Local Universe \citep{garnet02, tortora22}. This points to a scenario where Leo A experienced significant gas accretion, while stellar feedback and mass loss regulate its present-day chemical evolution.

\end{itemize}

\begin{acknowledgements} 
We thank the anonymous referee for their useful and professional comments which improved the quality of the paper. L.M. and A.A acknowledge support from ANID-FONDECYT Regular Project 1251809.
\end{acknowledgements}


% -------------------------------------------------------------------------------------
% -------------------------------------------------------------------------------------
% -------------------------------------------------------------------------------------
% -------------------------------------------------------------------------------------
% -------------------------------------------------------------------------------------
% -------------------------------------------------------------------------------------
% -------------------------------------------------------------------------------------
\begin{appendix}
\section{Mock slits in the VIMOS-IFU data cube}
\label{appendix:mock_slits}
\begin{figure}
\centering
\includegraphics[width=\hsize]{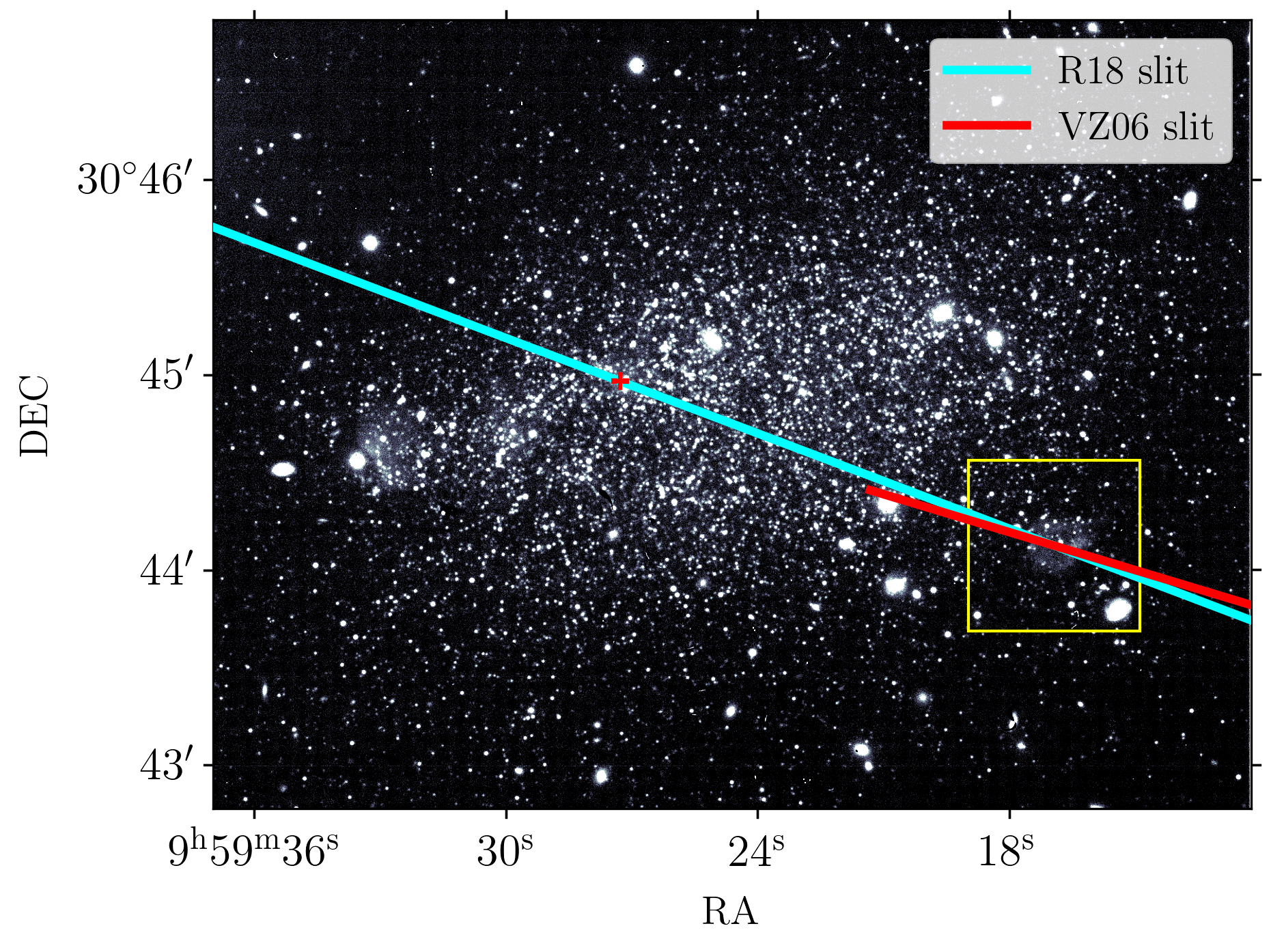}
\caption{Subaru Suprime-Cam H$\alpha$ frame from \citeauthor{stonkute14} (\citeyear{stonkute14}, \citeyear{stonkute19}). The yellow square is the VIMOS-IFU FOV. The cyan line represents the \citetalias{ruizescobedo18} slit position, which goes from \hii west (yellow square), to the \hii east (red cross). The red line represents the \citetalias{vanzee06} slit position.} 
\label{fig_slit_config}
\end{figure}

\begin{figure*}
\centering
\includegraphics[width=\hsize]{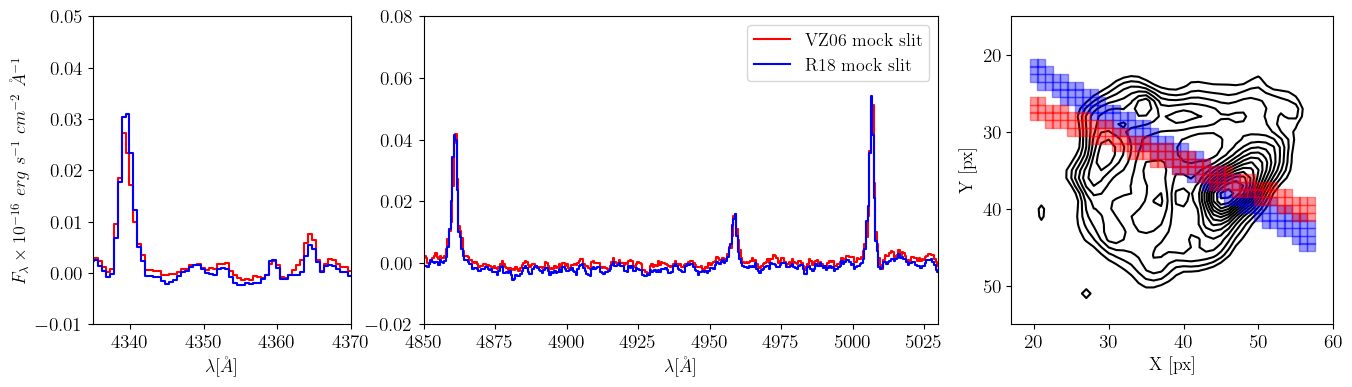}
\caption{Integrated spectrum of the Leo A \hii region using the \citetalias{vanzee06} and \citetalias{ruizescobedo18} mock slits, with red and blue colours, respectively. Left panel: spectral window showing H$\gamma$ and \oiiiauroral detection, from left to right, respectively. Middle panel: spectral window presenting the H$\beta$, and the $[\ion{O}{iii}]\lambda\lambda 4959,5007$ detection, from left to right, respectively. Right panel: spatial distribution of pixels selected to reproduce both mock slits in the \hii region. The grey contours represent the \hb emission of the nebula for reference.} 
\label{fig_slit_spectra}
\end{figure*}

\nolinenumbers
The $T_{e}$-based total oxygen abundances require the \oii emission line, which lies outside the VIMOS-IFU spectral range. Consequently, it is not possible to derive directly \ophp  from our data. To deal with this limitation, we simulated the slit positions and orientations of the long-slit observations of \citetalias{vanzee06} and \citetalias{ruizescobedo18}, allowing us to combine their \oii fluxes with our IFU results to compute total oxygen abundances.\\
\\
\citetalias{vanzee06} observed the \hii region, labelled $-101-052$, with a $2''$ wide slit centred at $\alpha = 9^{h}59^{m}17.2^{s}$ and $\delta = +30\degree 44' 07''$ (J2000), and position angle of $73\degree$. On the other hand, \citetalias{ruizescobedo18} employed a $7.4'$ long and $1.8''$ wide slit to cover two \hii regions of Leo A: the western \hii region studied here (\hii-west) and an eastern \hii region at $\alpha= 9^{h}59^{m}24.5^{s}; \ \delta= 30^{\circ} 44' 59''$ (J2000; see \citetalias[Figure 1]{ruizescobedo18}).\\
\\
The slit positions are shown in Figure \ref{fig_slit_config}, where the yellow square presents the VIMOS-IFU FOV (covering $\ion{H}{ii}$-west). The cyan line shows the \citetalias{ruizescobedo18} slit, and the red line the \citetalias{vanzee06} slit. The red cross marks the location of $\ion{H}{ii}$-east. A slight inclination difference between the two slit orientations is observed.\\
\\
Since the VIMOS spatial scale is $0.67''$ px$^{-1}$, we reproduced both slits as extractions of 3 px ($2.01''$) width on the IFU cube. The resulting integrated spectra are presented in Figure \ref{fig_slit_spectra}, showing the VZ06 mock slit in red and the R18 mock slit in blue (left and middle panels). The right panel shows the position of the mock slits over the \hb emission line map for reference (black contours). Both mock-slit spectra exhibit a detectable \oiiiauroral emission line, and present in general similar emission line fluxes with slight differences. The corresponding flux measurements and dust-corrected intensities are listed in Table \ref{table1} of Section \ref{detection}.

\section{Combination of long-slit and the VIMOS-IFU mock-slit spectrum for $T_{e}-$based metallicity estimations}
\label{appendix:te}

\begin{table*}[]
\centering
\caption{Electron temperatures, ionic and total oxygen abundances estimates for the Leo A \hii region.}

\begin{tabular}{cccccc}
\hline

                                & VZ06  & VZ06 VIMOS mock slit &  R18 & R18 VIMOS mock slit & VZ06 + R18  \\
                                & Eq. \ref{eq:no_combination_to}  & Eq. \ref{eq:combination_te} &  Eq. \ref{eq:no_combination_to} & Eq. \ref{eq:combination_te} & Eq. \ref{eq:combination_vz_r18}  \\

\hline \hline
$T_{e}[\ion{O}{iii}]$            &  --    &  $22693\pm 426$  &  --  &  $22332\pm 2047$ &  $22332\pm 2047$ \\
$T_{e}[\ion{O}{ii}]$           &  --    &  $18885 \pm 760$  &  --  & $18632 \pm 2348$ & $18632 \pm 2348$ \\
12+$\log(\mathrm{O^{+}/H})$     &   $7.11\pm0.06$   & -- & $7.26\pm 0.02$ &  -- & $7.11\pm 0.05$  \\
12+$\log(\mathrm{O^{++}/H})$    & $6.69\pm0.07$ &  $6.81\pm0.04$ &  $6.78\pm0.06$   & $6.84\pm0.08$ & $6.77 \pm 0.04$  \\
12+$\log(\mathrm{O/H})_{T_{e}}$ & $7.29\pm0.04$ & $7.29\pm0.06$ &  $7.38\pm0.05$   &   $7.40\pm0.10$  &  $7.28\pm0.04$ \\
\hline
   &  Eq. \ref{eq:r23_no_combination} & Eq. \ref{eq:r23_combination} & Eq. \ref{eq:r23_no_combination} & Eq. \ref{eq:r23_combination} & Eq. \ref{eq:r23_combination_vz_r18}       \\

12+$\log(\mathrm{O/H})_{R_{23}}$   & $7.43\pm0.09 $ & $7.45\pm0.07$ & $7.65\pm0.06$ & $7.65\pm0.07$     & $7.44 \pm 0.07$       \\
\hline

\end{tabular}
\label{table_appendix_mets}

\tablefoot{The second columns shows the estimates using the \citetalias{vanzee06} fluxes, applying dust correction described in Section \ref{detection}. The third column presents estimates using the VZ06 VIMOS mock slit (red colour in Figure \ref{fig_slit_spectra}). The fourth columns show the estimates using the \citetalias{ruizescobedo18} intensities. The fifth column presents the estimates using the R18 VIMOS mock slit (blue colour in Figue \ref{fig_slit_spectra}).}
\end{table*}

As mentioned in the main text, the \oii emission line is outside the VIMOS-IFU spectral range, so only \opphp can be measured directly. To derive total oxygen abundances, we therefore combined \oii fluxes to obtain \ophp from the long-slit spectra of \citetalias{vanzee06} and \citetalias{ruizescobedo18} with \opphp estimates from our VIMOS mock slits.

\subsection*{Procedure}
We estimated total oxygen abundances as follows:
\begin{itemize}

    \item (i): Extract \oii and \hb from \citetalias[Table 3]{vanzee06} to obtain \ophp.

    \item (ii): Compare \opphp derived from the VZ06 mock slit with that obtained using \citetalias[Table 3]{vanzee06} fluxes.

    \item (iii): Combine \ophp and \opphp to compute $T_{e}$-based metallicity. 
    
    \item (iv): Use \oii over \hb from \citetalias[Table 3]{vanzee06} to combine with \oiiistrongweak+\oiiistrong over \hb from their mock slit to compute $R_{23}$-based metallicities. Then, compare with $R_{23}$ metallicities using \citetalias{vanzee06} fluxes only.
    
    \item (v) Repeat (i), (ii), (iii), and (iv) with \citetalias[Table 3]{ruizescobedo18} fluxes and the R18 mock slit.

\end{itemize}

To ensure consistency, we applied our dust correction to the raw \citetalias{vanzee06} fluxes, since their correction assume Case B recombination with H$\gamma$/H$\beta$ = 0.474 at $T_{e}=15000$ K, and $n_{e}=100$ cm$^{-3}$, whereas we adopt H$\gamma$/H$\beta$ = 0.468 at $T_{e}=10000$ K, and $n_{e}=100$ cm$^{-3}$.

\subsection*{Combination of ionic abundances}
Total oxygen abundances were obtained by combining \ophp from \citetalias{vanzee06} and \citetalias{ruizescobedo18} with \opphp from the VIMOS mock slits, as follows
\begin{equation}
\label{eq:combination_te}
    12+\log\left(\mathrm{\frac{O}{H}}\right) = 12 + \log\left( \mathrm{\frac{O^{+}}{H^{+}}}\bigg\rvert_{\mathrm{VZ06\ or \ R18}} + \mathrm{\frac{O^{++}}{H^{+}}}\bigg\rvert_{\mathrm{VIMOS\ mock}} \right)
\end{equation}
For comparison, abundances were also derived using only the long-slit data, under $T_{e}$ derived from the VIMOS mock slits:
\begin{equation}
\label{eq:no_combination_to}
    12+\log\left(\mathrm{\frac{O}{H}}\right) = 12 + \log\left(\mathrm{\frac{O^{+}}{H^{+}}}+ \mathrm{\frac{O^{++}}{H^{+}}} \right)\bigg\rvert_{\mathrm{VZ06 \ or \ R18}}
\end{equation}
The reliability of the combination will be observed as consistency within the uncertainties between both equations, as well as the ionic \opphp abundances from the VIMOS mock slits and the ones derived from the long-slit datasets of VZ06 and R18.\\
\\
Temperatures were combined with getTemDen Pyneb module from the $[\ion{O}{iii}]\lambda\lambda4959,5007 / \lambda4363$ ratio, and $T_{e}[\ion{O}{ii}]$ from the relation of \citet{campbell86}. Ionic abundances were obtained with the getIonAbundance Pyneb module.

\subsection*{Results and discrepancies}
The values derived are listed in Table \ref{table_appendix_mets}. \citetalias{vanzee06} and its VIMOS mock slit are in agreement, returning $12+\log(\mathrm{O/H})\simeq7.29$ dex. \citetalias{ruizescobedo18} and their VIMOS mock slit are also in agreement within uncertainties, as $7.38\pm0.05$ and $7.40\pm0.10$, respectively. However, those are systematically higher than \citetalias{vanzee06} by $\sim0.1$ dex, despite their similar $T_{e}$ values ($\sim 22300$ K).\\
\\
Further exploration led to an analysis of ionic oxygen abundances. We found that the offset comes from \ophp estimates. The \citetalias{ruizescobedo18} \ophp estimates are higher by $\sim0.15$ dex. This may come from an inaccurate \oii emission line estimation, since the emission line falls in the region with the highest spectral noise \citepalias[Figure 4]{ruizescobedo18}, and the line is hard to detect. In contrast, \citetalias[Figure 2]{vanzee06} report a clear \oii detection despite the spectral noise in that region.\\
\\
Because \opphp agrees well between both datasets, to explore how the metallicity estimate given by \citetalias{ruizescobedo18} would change using a more accurate \oii detection, we combine \ophp from \citetalias{vanzee06} with \opphp from \citetalias{ruizescobedo18}, as:
\begin{equation}
\label{eq:combination_vz_r18}
    12+\log\left(\mathrm{\frac{O}{H}}\right) = 12 + \log\left( \mathrm{\frac{O^{+}}{H^{+}}}\bigg\rvert_{\mathrm{VZ06}} + \mathrm{\frac{O^{++}}{H^{+}}}\bigg\rvert_{\mathrm{R18}} \right)
\end{equation}
which return $12+\log(\mathrm{O/H})=7.28\pm0.04$, consistent with \citetalias{vanzee06} alone and our mock-slit combination.

\subsection*{$R_{23}$ consistency check}
As a final test, we computed the $R_{23}$ and its respective empirical metallicity using the calibration of \citet{kk04}. The index was constructed by combining the \oii and \hb from the long slit data with \oiiistrongweak, \oiiistrong, and \hb from the VIMOS mock slits, as follows:
\begin{equation}
\label{eq:r23_combination}
    R_{23} = \left( \frac{I(3727)}{H\beta}\right)_{\mathrm{VZ06\ or\ R18}} + \left( \frac{I(4959)+I(5007)}{H\beta}\right)_{\mathrm{VIMOS\ mock}}
\end{equation}
For comparison, the $R_{23}$-based oxygen abundances were obtained by using only long-slit fluxes of \citetalias{vanzee06} and \citetalias{ruizescobedo18}. The $R_{23}$ index is expressed as:
\begin{equation}
\label{eq:r23_no_combination}
    R_{23} = \left( \frac{I(3737)+I(4959)+I(5007)}{H\beta} \right)_{\mathrm{VZ06\ or\ R18}}
\end{equation}
The $R_{23}$-based metallicities derived for \citetalias{vanzee06} fluxes and their mock slit are $7.43\pm0.09$ and $7.45\pm 0.07$, respectively, showing consistency. The same is observed for \citetalias{ruizescobedo18} and their mock slit, as $7.65\pm 0.06$ and $7.65\pm0.07$, respectively. However, the \citetalias{ruizescobedo18} measure is $\sim 0.2$ dex higher than \citetalias{vanzee06}. Due to the discrepancies detected in the \ophp ionic abundances, we also constructed the $R_{23}$ index by combining \citetalias{vanzee06} and \citetalias{ruizescobedo18}, to explore how the $R_{23}$-based metallicity would change with an accurate \oii detection as follows:  
\begin{equation}
\label{eq:r23_combination_vz_r18}
    R_{23} = \left( \frac{I(3727)}{H\beta}\right)_{\mathrm{VZ06}} + \left( \frac{I(4959)+I(5007)}{H\beta}\right)_{\mathrm{R18}}
\end{equation}
which return a $R_{23}$-based metallicity of $7.44\pm 0.07$, in agreement with the \citetalias{vanzee06} and VZ06 mock slit results. This is also in agreement with the \citetalias{vanzee06} empirical ($\sim$7.48 dex) and the semiempirical ($7.44 \pm 0.10$ dex) estimates reported in their Table 6.

\section{Correction by temperature fluctuations}
\label{appendix:te_corrections}

The direct method is known to suffer from intrinsic biases \citep{kobulnicky96, pilyugin16}. These include (i) discrepancies between $T_{e}$ estimates from recombination and collisional lines, which can produce an \opp abundance difference up to three orders of magnitude \citep{esteban14}, and (ii) temperature fluctuations with non-homogeneous \hii regions. The latter cause overestimated $T_{e}$ determinations, since line emissivities scale as $e^{-h\nu/kT_{e}}$, and therefore lead to underestimated total oxygen abundances \citep{peimbert67}.\\
\\
As shown in the DESIRED sample of extragalactic HII regions \citep{mendez-delgado23b}, thermal stratification can introduce small-scale temperature inhomogeneities. In low-metallicity regimes, the cooling mechanisms are inefficient \citep{osterbrock06}, and these thermal gradients may introduce subtle $T_{e}$ variations, such as the observed in the SagDIG HII region \citep{andrade25}. However, they may remain below the resolution limit of IFU observations of this Leo A \hii region. Such variations, if present, would bias the total oxygen abundance determination with the direct method toward lower values \citep{peimbert17}.\\
\\
For this reason, although our measurements reveal no significant temperature fluctuations across the nebula, the presence of small-scale $T_{e}$ variations can not be entirely ruled out. Therefore, we include a correction for small-scale $T_{e}$ fluctuations. This correction does not imply that $T_{e}$ flucutations are detected in this \hii region, but instead provides a conservative and a physically motivated estimate of this systematic bias that should be taken into account when total oxygen abundances are determined with the direct method.\\
\\
Under the \citet{peimbert67} formalism, these are quantified by the electron-temperature root mean square parameter $t^{2}$, requiring $T_{e}$ estimates from multiple ionic species, and a mean electron temperature $T_{0}$. In principle, $T_{0}$ can be derived from $[\ion{N}{ii}]\lambda\lambda5755,6584$ emission lines \citep{mendez-delgado23}, but the VIMOS-IFU data do not cover the spectral region where $[\ion{N}{ii}]\lambda6584$ falls.\\
\\
We therefore adopt the empirical calibration of \citet{cameron23}, based on the RAMSES-RTZ simulations of an isolated dwarf galaxy \citep{katz22}, which is a similar case of Leo A. This method requires only the standard emission lines ($[\ion{O}{ii}]\lambda3727$, $[\ion{O}{iii}]\lambda\lambda\lambda 4363,4959,5007$) and defines a "line temperature", related to the emissivity of the emission line in a range of probed temperatures (see the scheme presented in their Figure 1). The line temperature, $T_{\mathrm{line}}$, is related to the auroral-based $T_{e}$ as $T_{line}=0.6 T_{ratio}+3258$ K, where $T_{ratio}$ is the $T_{e}$ derived from \oiiiauroral auroral line. \\
\\
For the Leo A \hii region, we obtain $T_{line}=16431\pm1041$ K, corresponding to a total oxygen abundance of $12+\log(\mathrm{O/H})\ = 7.46 \pm 0.09$. It is important to mention that this approach is based on just one isolated dwarf galaxy simulated, and the probed range of \citet{cameron23} goes from $\sim12000$K to $\sim17000$ K, and metallicities from 7.5 dex to 8.5 dex, being out of the range of Leo A. However, the corrected metallicity is $\sim0.17$ dex higher, consistent with the expected fluctuation levels of the \citet{cameron23} and also in line with the expected fluctuations observed in extragalactic HII regions \citep{mendez-delgado23b}. In addition, this correction does not affect the interpretations presented in Section \ref{discussionssss}. To apply calibrations that takes into account $T_{e}$ flucutations in dIrrs of the Local Group, those should be constructed for the metal-poor regime (\mets$\leq 7.5$ dex), based on both simulations and observations of spatially resolved extragalactic \hii regions.

\section{Classification between young MS stars and old stars in the Leo A \hii region}
\label{appendix:CMD}
Similar to the analysis presented in \citet{andrade25}, we classify the stellar populations in the Leo A \hii region FoV between young MS and old stars, since we aim to study a link between the stellar population and the ionised structure of the Leo A \hii region (Section \ref{hstcomparison}). We noted that the photometry acquired from the Hubble Source Catalog presents an offset of $\sim0.5$ mag in the $\mathrm{F814W}$ filter when it is compared with the \citet{cole07} photometry. So we apply this value as a correction.

\begin{figure}
\centering
\includegraphics[width=\hsize]{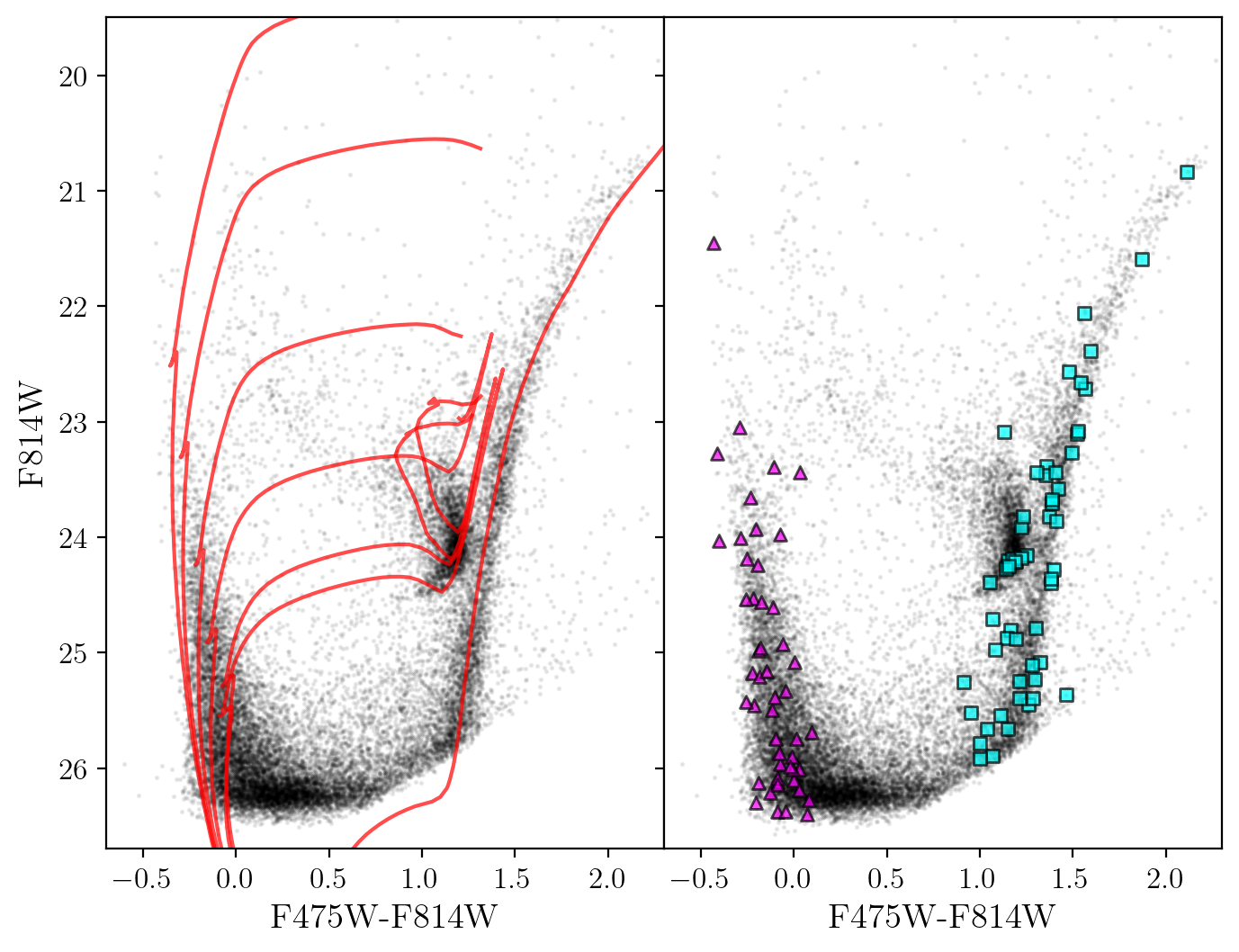}
\caption{HST/ACS CMD of the Leo A galaxy shown with black dots. Left panel: theoretical PARSEC isochrones (red curves) used to classify in the range of ages 50 Myr, 100 Myr, 250Myr, 500 Myr, 800 Myr, 1 Gyr, and 5 Gyr, adopting Z=0.0007, and $A_{814} = 0.039$. Right panel: selected young MS stars (magenta triangles) and old stars (cyan squares) in the Leo A \hii region FoV.} 
\label{fig6}
\end{figure}

To classify between those two stellar populations, we employed isochrones of the theoretical PARSEC v1.2S collection \citep{bressan12} and the Leo A HST/ACS CMD, adopting a metallicity of Z=0.007 (\citealt{stonkute19}, \citealt{urbaneja23}), extinction $A_{814} = 0.039$ and $A_{475}=0.078$ \citep{cole07}, and a distance of 800 Kpc (\citealt{dolphin02}, \citealt{bernard13}). Isochrones of ages of 50 Myr, 100 Myr, 250 Myr, 500 Myr, 800 Myr, 1 Gyr, and 5 Gyr, are presented as red curves in the left panel of Figure \ref{fig6}, from left to right, respectively. The stars belonging to the blue plume are compatible with ages $<500$ Myr, whereas the red clump stars are in line with ages between 800 Myr and 1 Gyr, and RGB stars seem to have ages of $\geq5$ Gyr. We proceed to classify the stars in the Leo A \ion{H}{ii} region FoV as young MS stars ($<500$ Myr) as those with $\mathrm{F475W - F814W} < 0.1$, and old stars ($\gtrsim 1$ Gyr), as those with $\mathrm{F475W - F814W} > 0.9$, presented as magenta triangles and cyan squares in the right panel of Figure \ref{fig6}.

\section{Behaviour of chemical evolution models}
\label{appendix:chem_models}

In Section \ref{chem_models_leoa}, we applied analytic chemical evolution models to explore the evolution of Leo A from 10 Gyr to the present. The equations 
\ref{eq:cb}, \ref{eq:acc}, \ref{eq:lb}, and \ref{eq:bt} describe the gas-phase metallicity, $Z_{g}$, as a function of the gas fraction, $\mu$, and the effective yield, \yeff.
\subsection*{Comparison of model at fixed yield}
Figure \ref{figchem0} shows the solution of the four models across the full range of gas fractions, $0<\mu<1$ for $y_{\mathrm{eff}}=0.02$ (solar yield). The closed-box and accretion models are identical at $\mu>0.5$, suggesting that galaxies which accrete more than half of their stellar mass in gas evolve effectively as closed boxes. By contrast, the leaky-box and gas-regulator models converge at $\mu>0.85$. At the extreme gas-rich regime ($\mu>0.9$), all four models produce nearly identical tracks, implying that extremely gas-dominated systems evolve as if no flows are present. The grey vertical bands represent the gas fractions of Leo A today ($\mu=0.68\pm0.05$) and 10 Gyr ago ($\mu=0.92\pm0.06$).

\subsection*{Behaviour of chemical evolution models with Leo A estimates}
\begin{figure}
\centering
\includegraphics[width=\hsize]{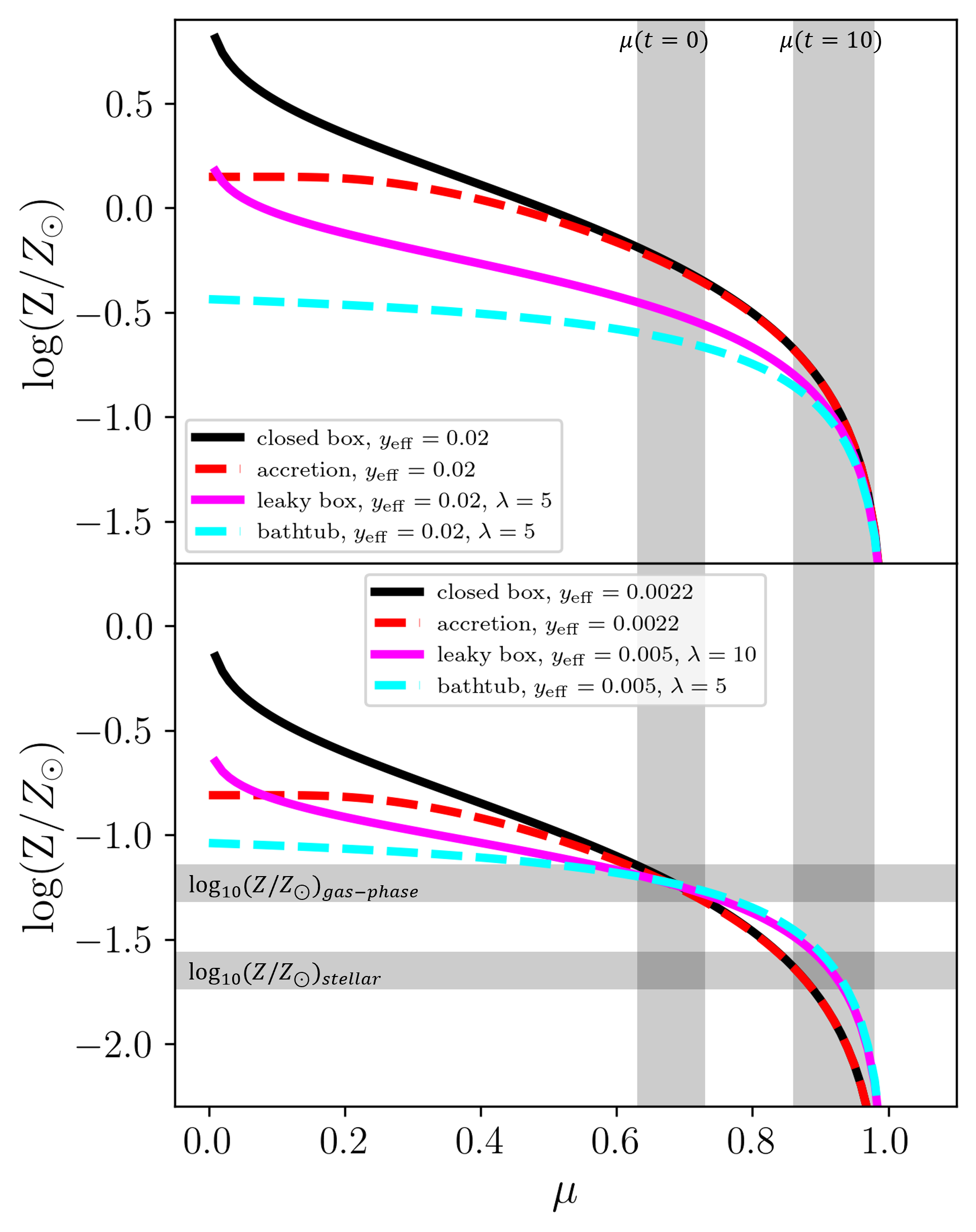}
\caption{The gas-phase metallicity as a function of the gas fraction, using the four chemical evolution models. Top panel: all models are reproduced by adopting solar metallicity as \yeff(0.02). The closed-box, accretion, leaky-box ($\lambda=5$), and gas-regulator ($\lambda=5$) models are shown with the black solid, red dashed, magenta solid, and cyan dashed curves, respectively. Bottom panel: same models, but using the estimated \yeff and $\lambda$ in Section \ref{chem_models_leoa} for Leo A. In both panels, the grey vertical bands show the estimated $\mu$ at 10 Gyr ago and the present day, whereas the horizontal bands show the stellar and gas-phase metallicities.} 
\label{figchem0}
\end{figure}

The bottom panel of Figure \ref{figchem0} shows the same comparison using the \yeff of Leo A derived by the Monte Carlo simulations. Here, grey horizontal bands are the gas-phase metallicity ($-1.65\pm0.09$ dex) and stellar metallicity ($-1.23\pm0.09$ dex; \citealt{kirby17}). The same qualitative behaviour is observed as in the top panel: closed-box and accretion models track each other closely, while leaky-box and gas-regulator models converge at high gas fraction but diverge at lower $\mu$.

\subsection*{The role of the effective yield}
\begin{figure*}
\centering
\includegraphics[width=\hsize]{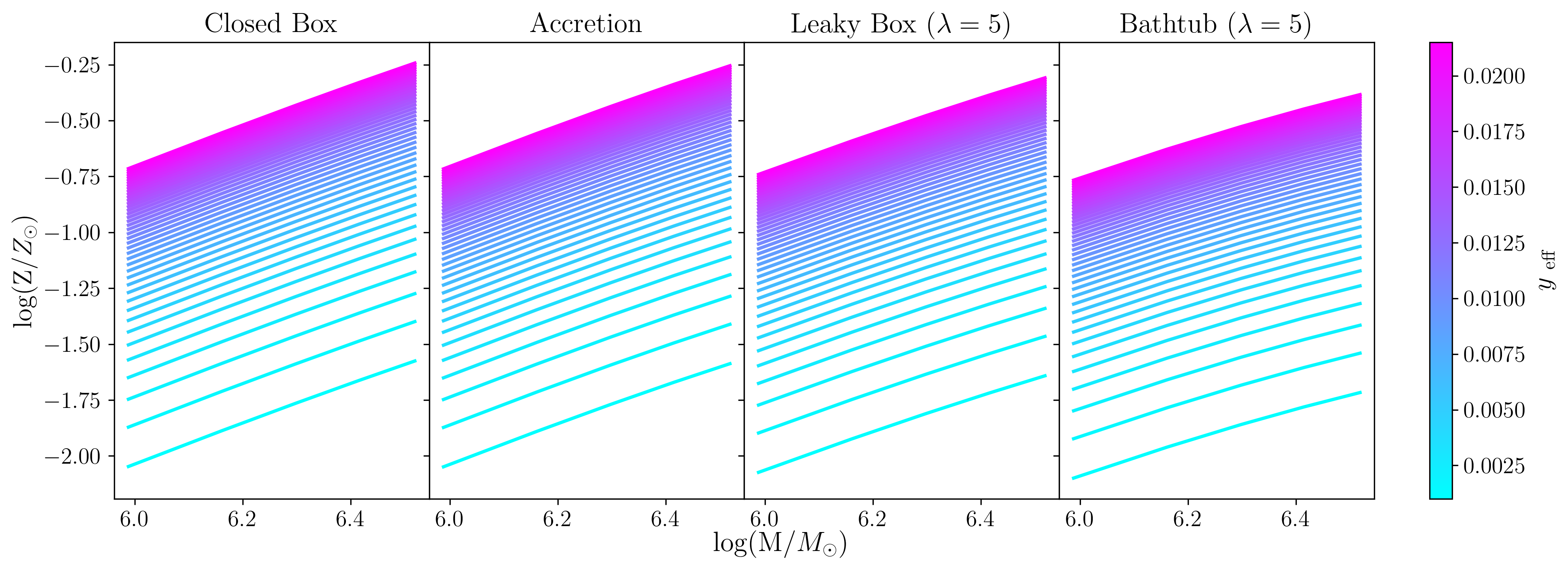}
\caption{Comparison of evolutionary tracks between the four chemical evolution models in the mass-metallicity plane. The colour code represents the \yeff of each track from 0.001 to 0.02 (solar). Closed-box, accretion, leaky box ($\lambda=5$), and bathtub ($\lambda=5$) are shown in the four panels, from left to right, respectively.} 
\label{figchem1}
\end{figure*}

The effect of \yeff is presented in Figure \ref{figchem1} for the four models. The colour code presents the \yeff evaluated, from 0.001 to 0.02. Varying \yeff shifts the intercept of the mass-metallicity curves: higher yields increase the metallicity at fixed gas fractions without altering the curve of the track in the mass-metallicity plane. 

\subsection*{The role of the mass-loading factor}
\begin{figure*}
\centering
\includegraphics[width=\hsize]{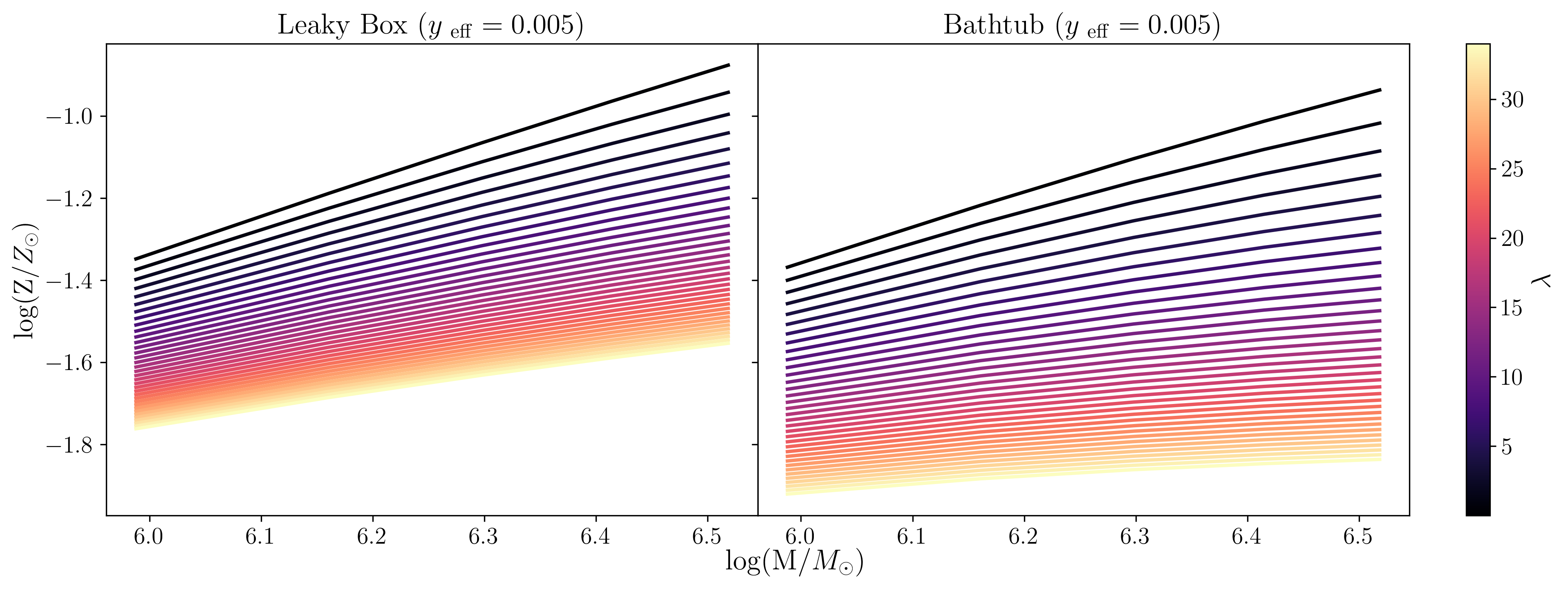}
\caption{Comparison of evolutionary tracks between the leaky-box model (left panel) and the bathtub model (right panel) under \yeff$=0.005$. The colour code represents the mass loading factor ($\lambda$) evaluated from 0 to 35.}

\label{figchem2}
\end{figure*}

Figure \ref{figchem2} shows the influence of the mass-loading factor, $\lambda$, in the leaky-box (left panel) and gas-regulator (right panel) models at \yeff$=0.005$, the value estimated for Leo A. The colour-code represents the $\lambda$ values evaluated, from 0 to 35. In this case, increasing $\lambda$ lowers the intercept and flattens the mass-metallicity curves, reflecting the suppression of enrichment by stronger outflows.

\subsection*{Summary}
These experiments clarify why closed-box and accretion models produce similar evolutionary tracks, and why leaky-box and gas-regulator models align under certain regimes. The convergence of all models at high gas fractions explains why gas-rich galaxies behave similarly using chemical evolution models, while the impact of \yeff and $\lambda$ sets the vertical normalisation and the slope of the mass-metallicity tracks for the chemical evolution of Leo A.

\end{appendix}

\end{document}